\documentclass[%
 reprint,
 amsmath,amssymb,
 aps,pra,
longbibliography
]{revtex4-1}

\usepackage{graphicx}
\usepackage{dcolumn}
\usepackage{bm}
\usepackage{hyperref}
\usepackage{xcolor}
\usepackage{amsmath, amssymb, amsfonts}
\usepackage{mathtools}
\usepackage{subfigure}

\definecolor{purple1}{rgb}{128,0,128}

\newcommand{\bea}{\begin{eqnarray}}
\newcommand{\ea}{\end{eqnarray}}
\newcommand{\ord}{{\cal O}}

\begin{document}


\title{Mesoscopics of half-quantum 
vortex pair deconfinement\\ in a trapped 
spin-one condensate}

\author{Seong-Ho Shinn}
\author{Uwe R. Fischer}
\affiliation{%
Seoul National University, Department of Physics and Astronomy, Center for Theoretical Physics, Seoul 08826, Korea
}%

\date{\today}

\begin{abstract}
Motivated by a recent experiment in an antiferromagnetic spin-1 Bose-Einstein condensate 
of ${}^{23} \textrm{Na}$ atoms, 
we study the energetical stability of a singly quantum vortex injected into the center of a quasi-two-dimensional gas with zero total spin against dissocation into a pair of half-quantum vortices. 
We find that the critical dissociation point of this confinement-deconfinement type phase transition can be expressed in terms of the ratio of density-density ($c_0$) and spin-spin ($c_2$) coupling constants.  
The transition of bound to unbound vortices, in particular, sensitively depends on (1) the ratio of system size ($R$) to density healing length ($\xi_d$), and (2) the trap potential. Specifically, the critical ratio $(c_2 / c_0)_{\textrm{cr}}$ increases when $R / \xi_d$ decreases, and  
is relatively larger in a harmonic trap than in a box trap. 
Dissociation is energetically generally favored for $c_2 / c_0 < (c_2 / c_0)_{\textrm{cr}}$, which as a corollary implies that vortex dissociation is observed as well for negative $c_2 < 0$, e.g., in a rubidium spin-1 BEC, whereas in a sodium spin-1 BEC ($c_2>0$) it is energetically blocked above the critical ratio 
$(c_2 / c_0)_{\textrm{cr}}$. 
Tuning the coupling ratio $c_2/c_0$ by using microwave control techniques, 
the dependence of the deconfinement phase transition on coupling parameters, density, 
and system size we predict, can be verified in experiments with ultracold spinor gases. 
\end{abstract}

\maketitle

\section{Introduction}

Topological defects occur in Nature after a 
symmetry-breaking second-order phase transition $G\rightarrow H$, where $G$ 
is the original, larger symmetry group,  and $H$ the remaining symmetry, whenever the 
homotopy group $\Pi_n (G/H)$ of the coset space $G/H$ is nontrivial~\cite{Kibble}. 
Of particular interest in the context of condensed matter 
is the first homotopy group $\Pi_1$, the so-called fundamental group, 
which leads to vortices, around whose center the neutral or charged 
liquid is entrained to circulate according to the rule of flux quantization~\cite{Mermin}.

A particularly intriguing species of vortex defects in the order parameter texture  
are half-quantum vortices (HQVs), examples of which are furnished 
in $^3\!$He~\cite{Grisha,Salomaa,Autti} 
and spin-triplet superconductors~\cite{Ivanov,Jang}, polariton condensates~\cite{HQV_observation_1st,SQV-HQV_dissoc_exp},  
as well as Bose-Einstein condensates (BECs)~\cite{Leonhardt,monopole_core_HQV_ring}.  
The first observations of vortices in 
BECs~\cite{1st_vortex_creation,Madison} ignited intense 
research on these hallmarks of superfluid behavior and the condensate phase.
The rich topology afforded by condensates with a multicomponent order parameter~\cite{vortices_in_BEC,Uedareview,UedaTopo,Jason,Ohmi} then led, in particular, to 
various studies regarding the implementation and dynamical properties of HQVs in 
BECs,~see,~e.g.~\cite{Zhou,JiHQV,Fetter,Shirley,Symes}.

HQVs also occur in the 
realm of a quantum simulation of high-energy physics. In the latter, they feature under the name Alice strings~\cite{Schwarz,Hindmarsh,GrishaBook}.
Furthermore, within the context of particle-topological defect duality,  
the quark confinement--deconfinement transition has been described by the dissociation of 
a singly quantized vortex (SQV) into HQVs in spinor BECs~\cite{HQV-Quark} and  in Rabi-coupled two-component condensates~\cite{vortex_pair_confine_Rabi}. 
%


A confinement-deconfinement phase transition from a SQV into a pair of HQVs was recently 
observed in experiment 
~\cite{ShinHQVExp}, where the SQV was initially 
injected into a quasi-two-dimensional (quasi-2D) sodium BEC 
in the polar phase, then transferred to the antiferromagnetic (AF) phase, in which vortex dissociation was observed.
The collisional dynamics of the HQVs was studied in~\cite{Collisional}; also cf.~the earlier experiments on quasi-2D skyrmions in \cite{Choi}.  
The present theoretical study is motivated by the experimental observation  
on HQV deconfinement in an ultracold spin-1 Bose gas, 
and aims at gaining a deeper understanding of the 
physical mechanism behind the dissociation process~\cite{Note}, as well as to stimulate further experiments
exploring the physics of HQV deconfinement. 


HQV interactions have previously been studied in two-component 
condensates in the infinite system limit~\cite{EtoInteraction,vortex_interaction}. 
Specifically, Ref.~\cite{EtoInteraction} found, 
under the condition $c_2 > 0$, at a critical point for which 
density-density ($c_0$) and spin-spin ($c_2$) coupling constants are equal, the vortex-antivortex
force changes from repulsive ($c_2 < c_0$) to attractive  ($c_2 > c_0$). 
Here, we consider a mesoscopically sized nonrotating spin-1 BEC in the AF phase, where the 
definition of mesoscopicity is afforded by the value of the ratio of system size $R$ to density-healing length $\xi_d$ ($\sim$ core size of SQV) being finite, to reveal the intricate interplay of density-density and spin-spin interactions in the vortex dissociation process.
We also demonstrate how the finite-size geometry helps us to reveal finer details of 
the physical origin of confinement versus deconfinement. 
In particular, we go beyond the study presented in~\cite{EtoInteraction} by  considering (1) a finite size, mesoscopic gas, (2) both box and harmonic trap potentials, and (3) both positive and negative
spin-spin coupling, $c_2 > 0$ and $c_2 < 0$, respectively. Our results are thus applicable to both sodium \cite{Stenger} respectively rubidium \cite{Chang} spin-1 condensates. 
Due to the definition of the density-density interaction coupling, $c_0 > 0$, for any spin-1 BEC~ in the thermodynamic limit to be stable \cite{Uedareview}. 
Denoting by $n$ the mean 3D number density of the spin-1 BEC, in a homogeneous gas, the ground state is the AF phase if $\left\vert p \right\vert < c_2 n $, $q < 0 $, and $c_2 > 0$~\cite{Uedareview}.
We then take as a working hypothesis below that the phase diagram of the finite-size spin-1 BEC  containing vortices is {\em qualitatively similar} (within our parameter regimes)   
to that of the infinite homogeneous spin-1 BEC.

To create vortices in the spin-1 BEC, one needs to disturb it by (sufficiently rapidly) rotating it~\cite{1st_vortex_creation} or by dragging a repulsive Gaussian laser beam through the gas~\cite{vortex_shedding}. In general, 
$E_{\textrm{ref}, j} < E_{V, j}$ where $E_{\textrm{ref}, j}$ is the total energy in a phase $j$ without vortex and $E_{V, j}$ is the energy in that same phase with vortices present.
However, due to topological conservation laws, once the vortex 
is created above the ground state, it will decay slowly to the ground state containing no vortex. 
Our aim is to derive the energetical stability against vortex dissociation from a SQV to a pair of HQVs 
in a quasi-2D spin-1 BEC; we assume as an axially symmetric initial condition that a SQV has been created at the center of the system. A BEC with two oppositely charged HQVs 
does not represent the ground state as well. Nevertheless, 
this configuration is potentially more stable than a BEC with a SQV, thus the dissociation from a SQV to two oppositely charged HQVs can be observable \cite{ShinHQVExp}.

\section{\label{general_setup}General Setup and Method} 

\subsection{\label{setup_ham_for_dim_reduc}Hamiltonian}

Two vortices with equal supercurrent winding numbers 
$q_n$ (see Def.~\eqref{defcharges} below)
rotate around each other with the center of mass being fixed, which has been verified for 
half-quantum vortices spin-1 BECs in~\cite{vortex_interaction}. To facilitate our calculations, we therefore use a co-rotating frame of two HQVs symmetrically placed at 
$\left( x, y \right) 
= \left( \pm D / 2 ,0 \right)$. 
Also, when we calculate the energy of the system with a SQV, for the same reason of retaining a sufficient degree of spatial symmetry,  we assume that the SQV is at the center of the system. 
Then, the Hamilton operator generally transforms according to 
$\hat H \rightarrow \hat H-\boldsymbol{\Omega} \cdot \hat{\boldsymbol{L}}$, with 
$\boldsymbol{\Omega} \cdot \hat{\boldsymbol{L}} = - i \hbar \Omega \left( \partial / \partial \varphi \right)$ where $\boldsymbol{\Omega} = \Omega \boldsymbol{e}_z$ is the angular velocity of the two co-rotating HQVs with respect to their center of mass (origin of coordinates), $\hat{\boldsymbol{L}}$ is angular momentum operator, and $\varphi$ the azimuthal angle. Therefore, due to cylindrical symmetry, the Hamiltonian with a SQV at the center remains invariant when going to the rotating frame. Using this fact, we may use the Hamilton operator for a nonrotating spin-1 BEC to calculate the energetical stability of the vortex dissociation process. The Hamilton operator is given by~\cite{Uedareview} 
\begin{eqnarray}
\hat{H} &=& \int d^3r \, \hat{\psi}^{\dagger}  \left\lbrack - \frac{\hbar^2}{2  M}  \nabla^2 + V_{\rm trap} \left( \boldsymbol{r} \right)  -  p  f_{z}  +  q  f^2_{z} \right\rbrack  \hat{\psi} 
\nonumber\\
&& \quad  + \frac{1}{2}  \int d^3 r\,
\left\lbrack 
c_{0} \left( \hat{\psi}^{\dagger}  \hat{\psi} \right)^2
 + 
c_{2} \left( \hat{\psi}^{\dagger}  \boldsymbol{f}  \hat{\psi} \right) \cdot \left( \hat{\psi}^{\dagger}  \boldsymbol{f} \hat{\psi} \right)
\right\rbrack, 
\nonumber\\
\label{Ham}
\end{eqnarray}
where $\hat \psi$ is the three-component spinor field operator and $V_{\rm trap}$ represents 
the scalar trapping potential, cf.~Eq.~\eqref{box_harmonic_text} below.
The coupling constants for density-density and spin-spin interactions are, respectively,  
$ c_0 = \left( g_0 + 2 g_2 \right) / 3 $ and $ c_2 = \left( g_2 - g_0 \right) / 3 $, where $ g_i = 4 \pi \hbar^2 a_B a_i / M $. Here, $ M $ is the mass of gas constituents, and $ a_{\mathcal{F}}$ 
is the $s$-wave scattering length of the spin-$ \mathcal{F}$ channel in units of the Bohr radius $ a_B $. Furthermore, $ \hbar \boldsymbol{f} $ is the spin-1 operator so that $ \left( f_z \right)_{m, m'} = m \delta_{m, m'} $ ($m, m' = -1, 0, 1$) where $ \delta_{m, m'} $ is the Kronecker delta. 
The operator of the total spin in $z$ direction is given by the integral of the 
spin density as \cite{Stamper-Kurn} 
\begin{eqnarray}
\hat S_z = \hbar \int d^3r \hat{\psi}^{\dagger} f_z \hat{\psi},  
\end{eqnarray}
and commutes with the 
Hamilton operator in Eq.~\eqref{Ham}, 
so that the integral of the magnetization, the total magnetic moment 
(obtained by multiplying $S_z$ with the magnetic moment of the spin-1 boson) 
is conserved. 
Also, $p$ 
denotes the linear Zeeman coefficient, and $q$ 
its quadratic counterpart.  

We operate with negative quadratic Zeeman shift, $q < 0$. 
According to~\eqref{Ham}, the quadratic Zeeman energy is  
$q \int d^3r ( \hat{\psi}^{\dagger}_1 \hat{\psi}_1 + \hat{\psi}^{\dagger}_{-1} \hat{\psi}_{-1} )$ where $\hat{\psi}_{m}$ is the component of $\hat{\psi}$ with magnetic quantum number $m_z = m$. 
Therefore, for  sufficiently large negative $q < 0$, and independent of the sign and magnitude of $c_2$, the total energy is lowered when $\hat{\psi}_0 \rightarrow 0$ because of the conservation of the norm
of $\hat\psi$. Assuming vanishing total spin, $\hat S_z = 0$, the ground state is then the AF phase, which leads to the wavefunction ansatz in Eq.~\eqref{imagansatz} below. 

Note that for negative spin-spin coupling $c_2<0$, ferromagnetic domains might potentially occur. 
However, within our subspace of 
vanishing total spin, we have verified that for the relatively small $|c_2/c_0|\sim \ord(1)$ 
we consider, the formation of ferromagnetic domains is energetically disfavored. 

Finally, because our focus is on spinor gases in the AF phase
and~\eqref{Ham} becomes independent of $p$ when $S_z = 0$, we fix $p = 0$ to 
facilitate our calculations, as $p$ is rendered dynamically irrelevant within a subspace of conserved vanishing total spin.


\subsection{\label{dim_reduc_main}Dimensional reduction}
We assume that a mean-field description of the quantum gas is applicable. Hence 
we replace the spinor field operator in \eqref{Ham} by its mean field, $\psi$. 

We consider the two types of trap potential experimentally commonly realized: harmonic and box traps.
To capture 
both within a single formula, 
the  scalar potential in \eqref{Ham} is assumed to be of  the form 
\begin{eqnarray}
V_{\rm trap} \left( \boldsymbol{r}, z \right) = 
\left\{
\begin{array}{ll}
\frac{1}{2} M \omega^2 \nu^2 r^2 + \frac{1}{2} M \omega^2_z z^2 & \textrm{if $ r < R $}
\\
\infty & \textrm{if $ r \ge R $}
\end{array}
\right. ,
\label{box_harmonic_text}
\end{eqnarray}
with $ \omega > 0 $ and $\nu \ge 0$. Here, $ \boldsymbol{r} $ is the position vector in the $x$-$y$ plane, and $ r \coloneqq \left\vert \boldsymbol{r} \right\vert $.
For this trap potential, we define scaled variables  according to
\bea 
\tau \coloneqq i \omega t,\qquad 
\tilde{r} \coloneqq \frac r l, \qquad  l = \sqrt{\frac{\hbar} {M  \omega} } .
\ea 
Here, $ \omega $ is a {\em scaling frequency} 
in the $x$-$y$ plane. 
For the box trap, we may set $ \omega $ to have an arbitrary value, and $ \nu = 0 $.
For a harmonic trap with $\omega_{\perp}$ the angular frequency of the harmonic trap in the $x$-$y$ plane, let $R_{\textrm{TF}}$ be the Thomas-Fermi (TF) radius  of the system in $x$-$y$ plane. We may then set $\nu = \omega_{\perp} / \omega$ and the system size is redefined as 
$ R = R_{\textrm{TF}} + \delta R$ where $\delta R$ is a nonnegative value introduced to aid the convergence of the numerical calculation. 
We therefore define $\tilde{R} \coloneqq R / l$.

For a quasi-2D spin-1 BEC, 
we employ the following ansatz for an AF mean-field wavefunction $ \psi \left( \boldsymbol{r},  t \right) $,
\begin{eqnarray}
&& \psi = 
\sqrt{\frac{N}{l^2  l_z  \sqrt{\pi}}}  e^{- i \left\{ \left( \omega_z / 2 \right) + \left( q / \hbar \right) \right\} t }  e^{- z^2 / \left\{ 2 \left( l_z \right)^2 \right\}} \tilde{F} \left( \tilde{r}, \varphi, \tau \right),
\nonumber\\
&& \tilde{F} \left( \tilde{r}, \varphi, \tau \right) = 
\left\lbrack 
\begin{array}{ccc}
- \tilde{f}_1 \left( \tilde{r}, \varphi, \tau \right) &
0 &
 \tilde{f}_{-1} \left( \tilde{r}, \varphi, \tau \right)
\end{array}\right\rbrack^T ,
\nonumber\\
\label{imagansatz}
\end{eqnarray}
where $ l_z =  \sqrt{\hbar / \left( M  \omega_z \right)} $ is the harmonic oscillator length along the $ z $ axis and $N$ the number of BEC atoms or molecules. The normalization condition is $\int d^3 r  \left\vert \psi \right\vert^2 = N$, and $\int d^3 r \, \psi^{\dagger} f_z \psi = 0$ because $S_z = 0$. Here, $\tilde{f}_{m} \left( \tilde{r}, \varphi, \tau \right)$ are complex functions.

Then, by assuming the transverse $z$ direction dynamics to be frozen to 
the harmonic oscillator ground state
\cite{Petrov,GPfortran} and correspondingly integrating it out, 
for $ r < R $ the following effective quasi-2D equations are obtained:
\begin{eqnarray}
-  \frac{\partial  \tilde{f}_{\pm 1}}{\partial \tau} = && 
\left\lbrack 
-  \frac{1}{2} \tilde{\nabla}^2_{2D} 
+ \frac{1}{2} \nu^2 \tilde{r}^2 
\right\rbrack  \tilde{f}_{\pm 1} 
\nonumber\\
&& \hspace*{-2em}+ N  C  
\left\lbrack
\left\{ 
\left( c_0' \pm c_2' \right)  \left\vert \tilde{f}_{1} \right\vert^2 
+\left( c_0' \mp c_2' \right)  \left\vert \tilde{f}_{-  1} \right\vert^2
\right\}  
\right\rbrack  \tilde{f}_{\pm 1} , 
\nonumber\\
\int_{0}^{2 \pi} d \varphi && \int_{0}^{\tilde{R}} d \tilde{r} \; \tilde{r} \left\vert \tilde{f}_{1} \right\vert^2 = \int_{0}^{2 \pi} d \varphi \int_{0}^{\tilde{R}} d \tilde{r} \; \tilde{r} \left\vert \tilde{f}_{- 1} \right\vert^2 = \frac{1}{2} .
\nonumber\\
\label{imaggpspin1_AF}
\end{eqnarray}
In Eq.~\eqref{imaggpspin1_AF}, $\tilde{\nabla}_{2D} \coloneqq \left( \partial / \partial \tilde{r} \right) \boldsymbol{e}_r + \left( 1 / \tilde{r} \right) \left( \partial / \partial \varphi \right) \boldsymbol{e}_{\varphi} $ where $C \coloneqq  2 \sqrt{2  \pi} \left( a_B / l_z \right)$, ${c_0'} \coloneqq  (a_0  +  2  a_2)/3$, and ${c_2'} \coloneqq  (a_2  -  a_0)/3$. For $ \tilde{r} \ge \tilde{R} $, $ \tilde{f}_m = 0 $.

By virtue of Eq.~\eqref{imaggpspin1_AF}, the density healing length $ \xi_d $ satisfies
\begin{eqnarray}
R / \xi_d = \sqrt{\frac{2}{\pi} N C c_0'} 
= \sqrt{\frac{2 M c_0}{\hbar^2} \frac{N}{\pi \left( l_z \sqrt{2 \pi} \right)}}  
\label{xid_AF}
\end{eqnarray}
and for $\nu \neq 0$, the TF radii $R_{\textrm{TF}}$ have the form 
\begin{equation}
R_{\textrm{TF}} = \frac{l}{\sqrt{\nu}} \sqrt[4]{\frac{4}{\pi} N C c_0'} = l_{\perp} \sqrt[4]{\frac{4}{\pi} N C c_0'},
\label{TFRad2D_AF}
\end{equation}
where $l_{\perp} = \sqrt{\hbar / \left( M \omega_{\perp} \right)} $ is harmonic oscillator length in the 
$x$-$y$ plane. 
Using Eq.~\eqref{xid_AF} for a quasi-2D system, we may define the volume of the BEC as $ \left( 4 \pi R^2 / 3 \right) \left(3 \sqrt{2 \pi} l_z / 4\right) $. Specific values of $\omega_{\perp}$ and $R$ for harmonically trapped gases will be introduced in section~\ref{harmonic_data}.

\subsection{\label{Ansatz_two_vortices} General Ansatz for up to Two Vortices}

We now (1) present our ansatz to calculate the wavefunction of the spin-1 quasi-2D BEC in the AF phase, with up to two vortices with opposite spin windings and (2) establish an energy criterion for
vortex dissociation, by defining the (scaled) energy difference Eq.~\eqref{scaled_E_diff} below. 
To this end, we first expound our general ansatz employed when quantum vortices are present in the system, 
which proved beneficial to reduce the computational time for both box (see section~\ref{box_data}) and  harmonic traps (see section~\ref{harmonic_data}).

\subsubsection{\label{vortex_ansatz_AF} Vortex ansatz in the AF phase}
We denote $q_n$ as the supercurrent winding number, and $q_s$ as the spin winding number~\cite{Uedareview,ShinHQVExp}, 
where, 
with the line integral taken around the central singularity in the vortex core, 
\begin{eqnarray}
\oint d \boldsymbol{l} \cdot \boldsymbol{v}_s = \frac{2\pi\hbar}{M} q_n, 
\quad 
\oint d \boldsymbol{\phi_s} = 2 \pi q_s .
\label{defcharges}
\end{eqnarray}
Here, $\phi_s$ is the azimuthal angle of spin orientation in the AF phase, $M$ is the mass of 
bosons, 
and $\boldsymbol{v}_s \coloneqq \left( \hbar / M \right) \textrm{Im} \left( \psi^{\dagger} \nabla \psi \right)$ is the superfluid velocity.

In order to consider a vortex pair containing a $ \left( q_n,  q_s \right) = \left( Q_n,  Q_s \right) $ vortex, whose core is at $ \left( \tilde{r},  \varphi \right) = \left( \tilde{r}_{F},  \varphi_{F} \right) $ and a $ \left( q_n,  q_s \right) = \left( Q_n,  -  Q_s \right) $ vortex, whose core is located at $ \left( \tilde{r}, \varphi \right) = \left( \tilde{r}_{G},  \varphi_{G} \right) $, we use the following ansatz:
\begin{eqnarray}
\tilde{f}_{\pm 1} \left( \tilde{r},  \varphi,  \tau \right)  && = 
\tilde{A}_{\pm 1} \left( \tilde{r},  \varphi,  \tau \right) \times
\nonumber\\
&&  \exp \left\lbrack i  \left\{
\Phi_{\pm 1} \left( \tilde{r},  \varphi \right)
 +  \Theta_{\pm 1} \left( \tilde{r},  \varphi \right)
 +  \tilde{B}_{\pm 1} \left( \tau \right)
\right\}
\right\rbrack,
\nonumber\\
\label{QVansatzpr}
\end{eqnarray}
where 
$ \Phi_{\pm 1} \left( \tilde{r},  \varphi \right) \coloneqq \left( Q_n  \mp  Q_s \right)  \phi_{F} \left( \tilde{r},  \varphi \right) $, 
$ \Theta_{\pm 1} \left( \tilde{r},  \varphi \right) \coloneqq \left( Q_n  \pm  Q_s \right)  \phi_{G} \left( \tilde{r},  \varphi \right) $,  and
\begin{eqnarray}
\cos \phi_{F} \left( \tilde{r},  \varphi \right) &\coloneqq&  
\frac{\tilde{r} \cos \varphi - \tilde{r}_F \cos \varphi_F}{\sqrt{\tilde{r}^2 + \left( \tilde{r}_{F} \right)^2 - 2  \tilde{r}  \tilde{r}_F \cos \left( \varphi - \varphi_F \right)}}, \nonumber\\
\cos \phi_{G} \left( \tilde{r},  \varphi \right) &\coloneqq&  
\frac{\tilde{r} \cos \varphi - \tilde{r}_G \cos \varphi_G}{\sqrt{\tilde{r}^2 + \left( \tilde{r}_{G} \right)^2 - 2  \tilde{r}  \tilde{r}_G \cos \left( \varphi - \varphi_G \right)}}.
\nonumber\\
\end{eqnarray}
Here, $ \tilde{A}_{\pm 1} $ and $ \tilde{B}_{\pm 1} $ are some real functions. Due to the single-valuedness of the wavefunction, $Q_n \pm Q_s$ should be integer.
Then, the superfluid velocity $ \boldsymbol{V}_s $ becomes
\begin{eqnarray}
\boldsymbol{V}_s
&& = \frac{\hbar}{M l} \frac{\displaystyle \sum_{m = \pm 1} \left\{ 
\tilde{\nabla}_{2D} \left( \Phi_{m} + \Theta_{m} \right) 
\right\} \tilde{A}^2_{m}}{\displaystyle \sum_{m' = \pm 1} \tilde{A}^2_{m'}}.
\label{supvel_ansatz}
\end{eqnarray}
For the present system with radius $ R $, let $ \Phi'_{\pm 1} $ be the phase of the image vortex of the $ \left( Q_n, Q_s \right) $ vortex at $ \left( \tilde{r}, \varphi \right) = \left( \tilde{r}_F, \varphi_F \right) $. In order to make the radial component of superfluid velocity vanish at the boundary, one imposes  
\begin{eqnarray}
&& \Phi'_{\pm 1} \left( \tilde{r},  \varphi \right) = - \left( Q_n \mp Q_s \right) \phi'_F \left( \tilde{r},  \varphi \right),
\nonumber\\
&& \cos \phi'_{F} \left( \tilde{r},  \varphi \right) \coloneqq  
\frac{\tilde{r} \cos \varphi -\tilde{r}'_F \cos \varphi_F}{\sqrt{\tilde{r}^2 + \left( \tilde{r}'_{F} \right)^2 - 2  \tilde{r}  \tilde{r}'_F  \cos \left( \varphi - \varphi_F \right)}},
\nonumber\\
\end{eqnarray} 
where $ \tilde{r}'_F = \tilde{R}^2 / \tilde{r}_F $ if $ \tilde{r}_F \neq 0 $. Observe that when $ \tilde{r}_F = 0 $, the $ \left( Q_n, Q_s \right) $ vortex is at the center of the system. Hence there is no radial component of superfluid velocity at the boundary of the system with radius $R$ and no image vortex is required to satisfy the boundary conditions. 
Likewise, we define $ \Theta'_{\pm 1} $ as the phase of the image vortex of the $ \left( Q_n, - Q_s \right) $ vortex at $ \left( \tilde{r}, \varphi \right) = \left( \tilde{r}_G, \varphi_G \right)$. 

When there is only one vortex with $ \left( q_n,  q_s \right) = \left( Q_n,  Q_s \right) $ at the center, $ \phi_G = \phi'_F = \phi'_G = \Theta_{\pm 1} = \Phi'_{\pm 1} = \Theta'_{\pm 1} = 0 $. When there is only one vortex with $ \left( q_n,  q_s \right) = \left( Q_n,  Q_s \right) $ off the center, $ \phi_G = \phi'_G = \Theta_{\pm 1} = \Theta'_{\pm 1} = 0 $. Finally, if there is no vortex, $ \phi_F = \phi_G = \phi'_F = \phi'_G = \Phi_{\pm 1} = \Theta_{\pm 1} = \Phi'_{\pm 1} = \Theta'_{\pm 1} = 0 $.

Employing the ansatz Eq.~\eqref{QVansatzpr}, the first line of 
Eq.~\eqref{imaggpspin1_AF} takes the form 
\begin{widetext}
\begin{eqnarray}
- \frac{\partial \tilde{A}_{\pm 1}}{\partial \tau} && = 
\left\lbrack
- \frac{1}{2} \tilde{\nabla}^2_{2D} 
+ \frac{1}{2} \left\vert \tilde{\nabla}_{2D} \left( \Phi_{\pm 1} + \Phi'_{\pm 1} + \Theta_{\pm 1} + \Theta'_{\pm 1} \right) \right\vert^2 
+ H_{\pm 1} 
\right\rbrack \tilde{A}_{\pm 1} ,
\nonumber\\
\tilde{A}_{\pm 1} \frac{\partial \tilde{B}_{\pm 1}}{\partial \tau} && = 
\left\lbrack
\left\{ \tilde{\nabla}_{2D} \left( \Phi_{\pm 1} + \Phi'_{\pm 1} + \Theta_{\pm 1} + \Theta'_{\pm 1} \right) \right\} \cdot \nabla_{2D} 
+ \frac{1}{2} \left\{ \tilde{\nabla}^2_{2D} \left( \Phi_{\pm 1} + \Phi'_{\pm 1} + \Theta_{\pm 1} + \Theta'_{\pm 1} \right) \right\} 
\right\rbrack \tilde{A}_{\pm 1} ,
\nonumber\\
&& \quad \textrm{ where } 
H_{\pm 1} \coloneqq 
\frac{1}{2} \nu^2 \tilde{r}^2 
+ N C \left\{
\left( c_0' \pm c_2' \right) \tilde{A}^2_{1}
+ \left( c_0' \mp c_2' \right) \tilde{A}^2_{-1}
\right\}
\textrm{ and }
\int_{0}^{2 \pi} d \varphi \int_{0}^{\tilde{R}} d \tilde{r} \; \tilde{r} \tilde{A}^2_{\pm 1} = \frac{1}{2} ,
\label{meanprdiff_AF}
\end{eqnarray}
\begin{eqnarray}
\mbox{and}\quad \frac EN = && \frac{\hbar \omega}2 \int_{0}^{2 \pi} d \varphi \int_{0}^{\tilde{R}} d \tilde{r} \; \tilde{r} 
\sum_{m = \pm 1}
\left\lbrack
\left( \tilde{\nabla}_{2D} \tilde{A}_{m} \right)^2 
+ \left\{
\left\vert \tilde{\nabla}_{2D} \left( \Phi_{\pm 1} + \Phi'_{\pm 1} + \Theta_{\pm 1} + \Theta'_{\pm 1} \right) \right\vert^2 
+ \nu^2 \tilde{r}^2
\right\} \tilde{A}^2_{m}
\right\rbrack 
\nonumber\\
&& + \frac{\hbar \omega}2 \int_{0}^{2 \pi} d \varphi \int_{0}^{\tilde{R}} d \tilde{r} \; \tilde{r} 
N C \left\lbrack
c_0' \left( \tilde{A}^2_{1} + \tilde{A}^2_{-1} \right)^2
+ c_2' \left( \tilde{A}^2_{1} - \tilde{A}^2_{-1} \right)^2
\right\rbrack
+ q + \frac{\hbar \omega_z}4 ,
\label{energy_AF}
\end{eqnarray}
\end{widetext}
where $E$ is the total energy of the system in the AF phase, according to 
Eqs.~\eqref{imaggpspin1_AF} and~\eqref{QVansatzpr}. Since the squared amplitude fulfills 
$ \left( l^2 l_z \sqrt{\pi} \right) \left\vert \psi_{\pm 1} \left( x, y, z, t \right) \right\vert^2 = N e^{- \left( z / l_z \right)^2} \left\vert \tilde{f}_{\pm 1} \left( \tilde{x}, \tilde{y}, \tau \right) \right\vert^2 = N e^{- \left( z / l_z \right)^2} \left\vert \tilde{A}_{\pm 1} \left( \tilde{x}, \tilde{y}, \tau \right) \right\vert^2$, $ \tilde{B}_{\pm 1} $ does not affect the density profile of the gas. Also, it does not affect the superfluid velocity, see~\eqref{supvel_ansatz} and the Hamiltonian~\eqref{Ham}. We therefore omit the $ \tilde{B}_{\pm 1} $ term in what follows. 

\subsubsection{\label{def_scaled_energy_diff}Definition of scaled energy difference}

Let $ E_{\textrm{ref}} $ be the total energy without vortex, $ E_S $ be that  with a SQV at the center, and $ E_H $ be that  with two oppositely charged HQVs, each at 
$ \left( x, y \right) = \left( \pm D / 2,0 \right) $. According to Eq.~\eqref{energy_AF}, as $N q + N \hbar \omega_z / 4 $ does not depend on $\tilde{A}_{\pm 1}$, when we assess  the energetical stability against vortex dissociation 
we examine the behavior of the (scaled) energy difference 
\begin{equation}
\Delta E_{H,S}\coloneqq 
\left( \tilde{E}_H - \tilde{E}_S \right) / \tilde{E}_{\textrm{ref}},
\label{scaled_E_diff}
\end{equation} 
where the shifted energies 
$\tilde{E}_{\textrm{ref}} \coloneqq E_{\textrm{ref}} - N q - N \hbar \omega_z / 4  $, 
$\tilde{E}_S \coloneqq E_S - N q - N \hbar \omega_z / 4  $, 
and $\tilde{E}_H \coloneqq E_H - N q - N \hbar \omega_z / 4  $. 
Hence, the exact value of $q$ is immaterial, as it does not appear in Eq.~\eqref{meanprdiff_AF} and in the (scaled) energy difference defined above. 


\subsection{Numerical Method}

The numerical method to solve Eq.~\eqref{meanprdiff_AF} is based on Refs.~
\cite{GP_numerical_methods_spin_1,GP_numerical_spin_1_method_derivations}, which employ a  second-order splitting method (so-called Strang splitting)~\cite{Strang}, to separate linear and nonlinear terms in the GP (Gross-Pitaevski\v\i) equation. 
In order to ensure that $S_z$ is conserved, a projection constant was introduced.
As $ \psi_0 = 0 $, the projection constant for $ \psi_{\pm 1} $ becomes the normalization factor of $ \psi_{\pm 1}$. In \cite{GPfortran,GPC}, a similar procedure was used to solve the {\em scalar} GP equation. 
 
A graphics processing unit (GPU) has thousands of cores which can perform a parallel computation only if one uses specific computing language. Supercomputers are capable of performing parallel computation, but one needs to assign the workloads to nodes which are parts of the supercomputer. 
Therefore, to reduce the total computing time by using parallel processing, 
we wrote two codes: one with OpenCL to use a GPU of AMD, and another code using a 
message passing interface (MPI) to take advantage of  a supercomputer for solving Eq.~\eqref{meanprdiff_AF}.

We studied 
gases trapped in box and harmonic trap potentials, whose form is given in Eq.~\eqref{box_harmonic_text},  fixing $ \omega_z / 2\pi = 400\, \textrm{Hz} $ (as employed in the sodium experiments \cite{ShinHQVExp}). 
The mass $M$ used is for $ {}^{23} \textrm{Na} $ atoms and $ c_0' $ is fixed to be $51.1$, which corresponds to the bare $ c_0' $ value of $ {}^{23} \textrm{Na} $~\cite{Uedareview}. 
We limited the pair size to $ D < 2 R $ where $ D $ is the distance between the two oppositely charged HQVs, because, naturally, {\em physical} vortex cores are located within the gas cloud. 
According to Eq.~\eqref{xid_AF}, $R / \xi_d = \sqrt{2 N C c_0' / \pi} = 0.0907 \sqrt{N}$. Note that $R / \xi_d$ does not depend on $\omega_{\perp}$. 
Therefore, once we set $R / \xi_d$, it is not necessary  to change $N$ irrespective of 
whether the trap is a box or harmonic one. 
Table~\ref{N_R_xi_d} 
displays the set of  $R / \xi_d$ and $N$ values used in this paper, together with the in-plane  
trap frequency $\omega_\perp$. 

\begin{table}[b]
\caption{\label{N_R_xi_d}
Employed parameter values $R / \xi_d$, $N$, and $\omega_{\perp}$, 
determined from Eq.~\eqref{xid_AF} and for a quasi-2D spin-1 gas of $ {}^{23} \textrm{Na} $ atoms.
The relation between $R / \xi_d$ and $N$ is identical whether the trap is box or harmonic due to the ansatz Eq.~\eqref{box_harmonic_text}. The vertical trap frequency $ \omega_z /2 \pi $ is thoughout fixed to be $ 400\, \textrm{Hz} $.
To test the code accuracy of our code, we also ran a sample calculation for  $R/\xi_d=4$ with 
$\omega_{\perp} / 2 \pi  = 5\, \textrm{Hz}$. 
}
\begin{ruledtabular}
\begin{tabular}{ccc}
$R / \xi_d$ & $N$ & $\omega_{\perp} / 2 \pi $ (Hz) \\
\hline
4 & 1943 & 0 \\
4 & 1943 & 5 \\
4 & 1943 & 20 \\
8 & 7774 & 0 \\
8 & 7774 & 20 \\
\end{tabular}
\end{ruledtabular}
\end{table}

We posit that $\tilde{A}_{\pm 1} \left( \tau = \tau_1 \right)$ is the ground state solution of Eq.~\eqref{meanprdiff_AF} if it satisfies the inequality 
$\displaystyle \int d^2 \tilde{r} \sum_{m = \pm 1} \left\vert \tilde{A}_{m} \left( \tau = \tau_1 \right) - \tilde{A}_{m} \left( \tau = \tau_0 \right) \right\vert^2 < \epsilon$, where $\tau_1 = \tau_0 + d \tau$ and $\epsilon$ is some small positive number which determines the convergence of the solution of Eq.~\eqref{meanprdiff_AF}; $d \tau >0 $ represents the imaginary time step size.
Then, since our ground state criterion in Eq.~\eqref{meanprdiff_AF} includes spatial integration, as $\tilde{R}$ decreases, $\epsilon$ should be decreased as well so that our ground state criterion 
is consistent independent of $\tilde{R}$.
In other words, we consider $\tilde{A}_{\pm 1} \left( \tau = \tau_1 \right)$ to be the ground state solution if it satisfies the following inequality,  
$\frac{1}{\pi \tilde{R}^2}  \int_{0}^{2 \pi} d \varphi \int_{0}^{\tilde{R}} d \tilde{r} \; \tilde{r} \sum_{m = \pm 1} \left\vert \tilde{A}_{m} \left( \tau = \tau_1 \right) - \tilde{A}_{m} \left( \tau = \tau_0 \right) \right\vert^2$  
$< \tilde{\epsilon}$.
Here, $\tilde{\epsilon}$ is another small number, set to be of the order of the typical machine precision,  
$\tilde\epsilon = \ord(10^{-14})$.

Defining $\Delta \tilde{r} > 0$ as the step size of $\tilde{r}$, and $ \Delta \varphi > 0$ as the 
step size of $\varphi$, when $\xi_d$ decreases, both 
$\Delta \tilde{r}$ and $ \Delta \varphi $ 
are required to remain small, because the numerical solutions of Eq.~\eqref{meanprdiff_AF} diverge if $ \Delta \tilde{r} \ge \textrm{Min} \left( \xi_d, \xi_s \right) / l$ or $ R \Delta \varphi \ge \textrm{Min} \left( \xi_d, \xi_s \right) $, where $\textrm{Min} \left( a, b \right) = a $ if $a \le b$ and $b$ if $a > b$, 
Here, $\xi_s$ satisfies $R / \xi_s = \sqrt{2 N C \left\vert c_2' \right\vert / \pi}$ by generalizing Eq.~\eqref{xid_AF} with~\cite{SQV_energetic_stability_spin_1}.

Given our numerical resources, we can study system sizes up to $R/\xi_d=8$. 
In addition, when limiting the numerical calculation time, there is a restriction on $c_2 / c_0$:
When $c_0 > 0$, Eq.~\eqref{meanprdiff_AF} effectively decouples when $c_2 = c_0$, whereas the coupling of $\tilde{A}_1$ and $\tilde{A}_{-1}$ increases as $c_2 < c_0$ or $c_2 > c_0$. Therefore, it
is more time-consuming to solve Eq.~\eqref{meanprdiff_AF} for $c_2 / c_0 \neq 1$.
Hence, limit ourselves to $-0.5 \le c_2 / c_0 \le 2$ given the resources.

\section{\label{cdct_results} Deconfinement of half-quantum vortices}

\subsection{\label{box_data} Box Traps}

\begin{figure} [b]
\centering
\includegraphics[width=0.244\textwidth]{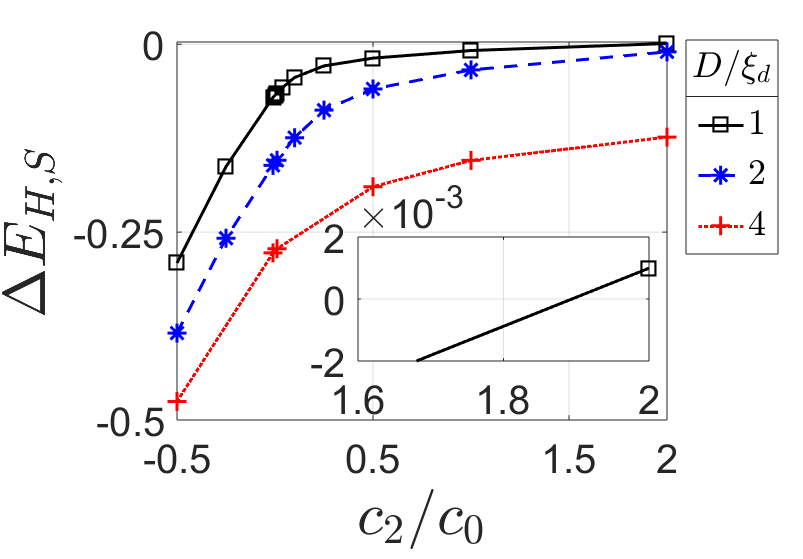}
\includegraphics[width=0.226\textwidth]{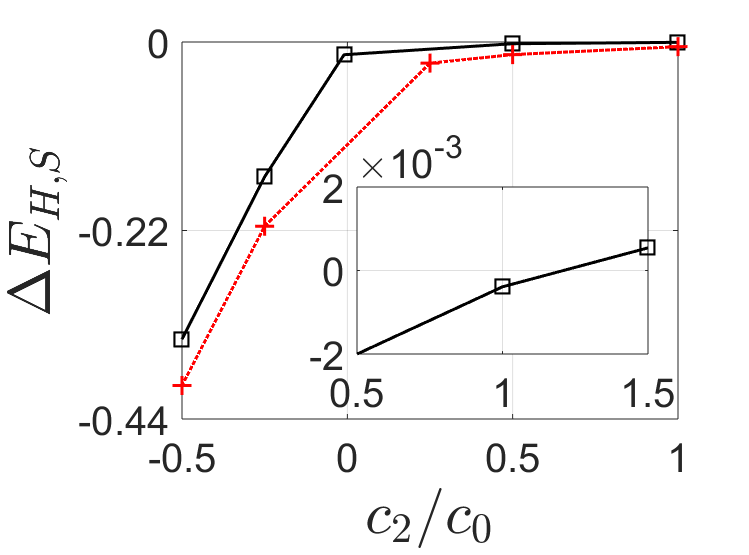}
\caption{
\emph{Scaled energy difference $\Delta E_{H,S}$ as a function of $ c_2 / c_0 $} for (a) $R / \xi_d = 4$, (b) $R / \xi_d = 8$
with box trap. Black dots are for $D / \xi_d = 1$, blue dots are for $D / \xi_d = 2$, and red dots are for $D / \xi_d = 4$, indicating how the  energetical stability against vortex dissociation depends on $c_2 / c_0$. The force between two oppositely charged HQVs are repulsive for every $c_2 / c_0$ region in this figure. 
The inset shows the critical crossing of $c_2 / c_0$  for $D/ \xi_d = 1$; 
horizontal and vertical axis are identical to the main plot. 
}
\label{fig2}
\end{figure}

When we consider box traps, $\nu = 0$ in Eq.~\eqref{meanprdiff_AF}. Also, there is no constraint imposed on $R$ like for harmonic traps: Once $N$ is determined, $R / \xi_d$ is automatically determined by Eq.~\eqref{xid_AF}. Similarly, $R / \xi_s$ is also determined by $c_2$.  
Note that our ansatz, Eqs.~\eqref{imagansatz} and~\eqref{meanprdiff_AF}, is valid for $0 \le \tilde{r} < \tilde{R}$ [see Eq.~\eqref{box_harmonic_text}], where $l$ is an harmonic oscillator type length scale  
which depends on the ``scaling frequency" $\omega$, as introduced below Eq.~\eqref{box_harmonic_text}. 
Therefore, to assess the energetics of vortex dissociation for box trap potentials, we may select one specific value of $R$. For example, suppose that one has solved Eq.~\eqref{meanprdiff_AF} for the system (Sys1) with box trap potential and 
$\tilde{R} = \tilde{R}_{\textrm{Sys1}}$ be the scaled system size. 
Let  $E - Nq - N \hbar \omega_z / 4 $ for that system be 
$\tilde{E}_{\textrm{Sys1}}$. 
Then, using Eq.~\eqref{energy_AF}, one finds for system Sys2 an energy expression equivalent to changing the spatial size of Sys1 by a factor
$\alpha$, 
$\tilde{E}_{\textrm{Sys2}} = \tilde{E}_{\textrm{Sys1}} / \alpha^2$. 
Using this feature, to reduce computing time, we therefore set 
$ \omega / 2 \pi = 5 \,\textrm{Hz} $  
to scale lengths in units of $ l = 9.37 \mu m $, and set $ \tilde{R} = 0.2 $.

\begin{figure*}[t]
\centering
\subfigure{
\includegraphics[width=0.49\linewidth]{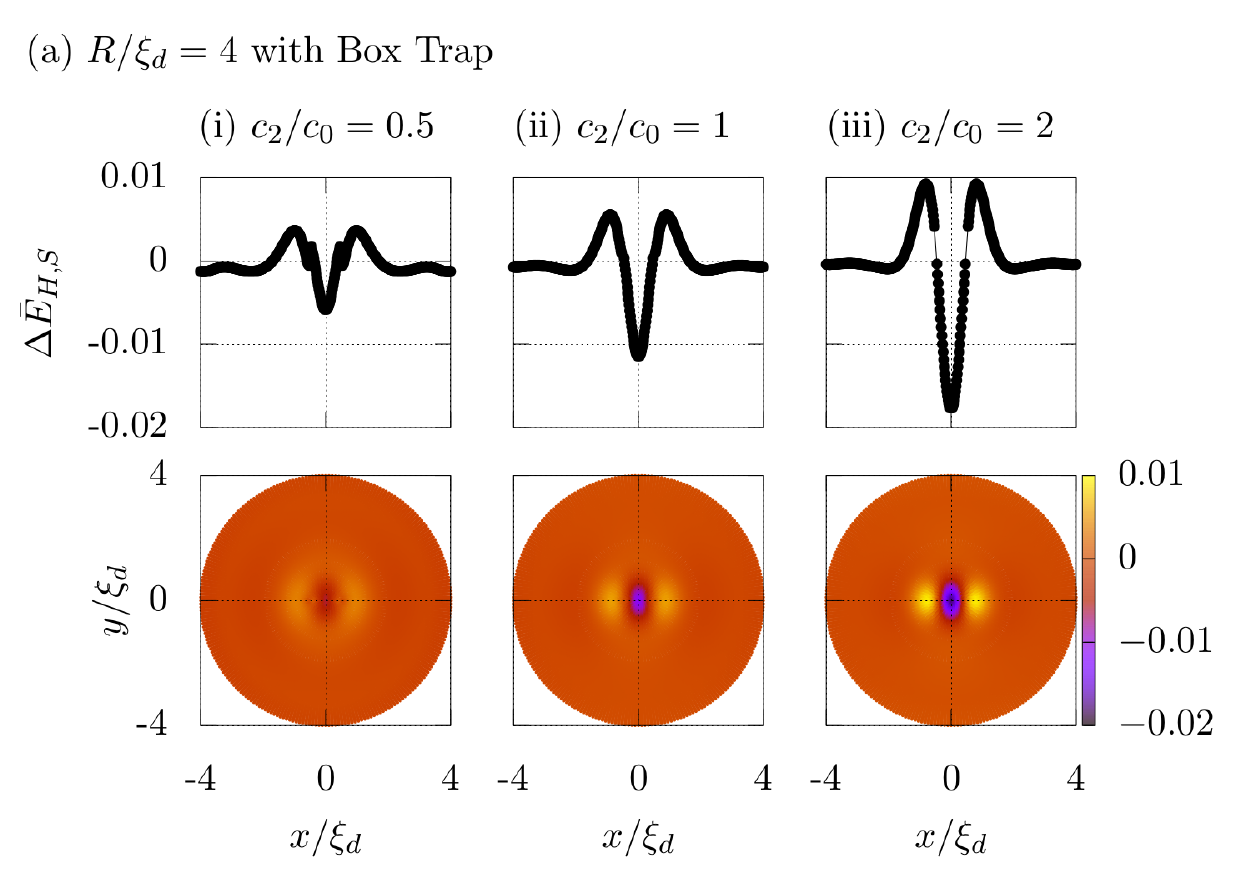}
\label{nt_R4_var_c2}
}
\hspace*{-1em}
\subfigure{
\includegraphics[width=0.49\linewidth]{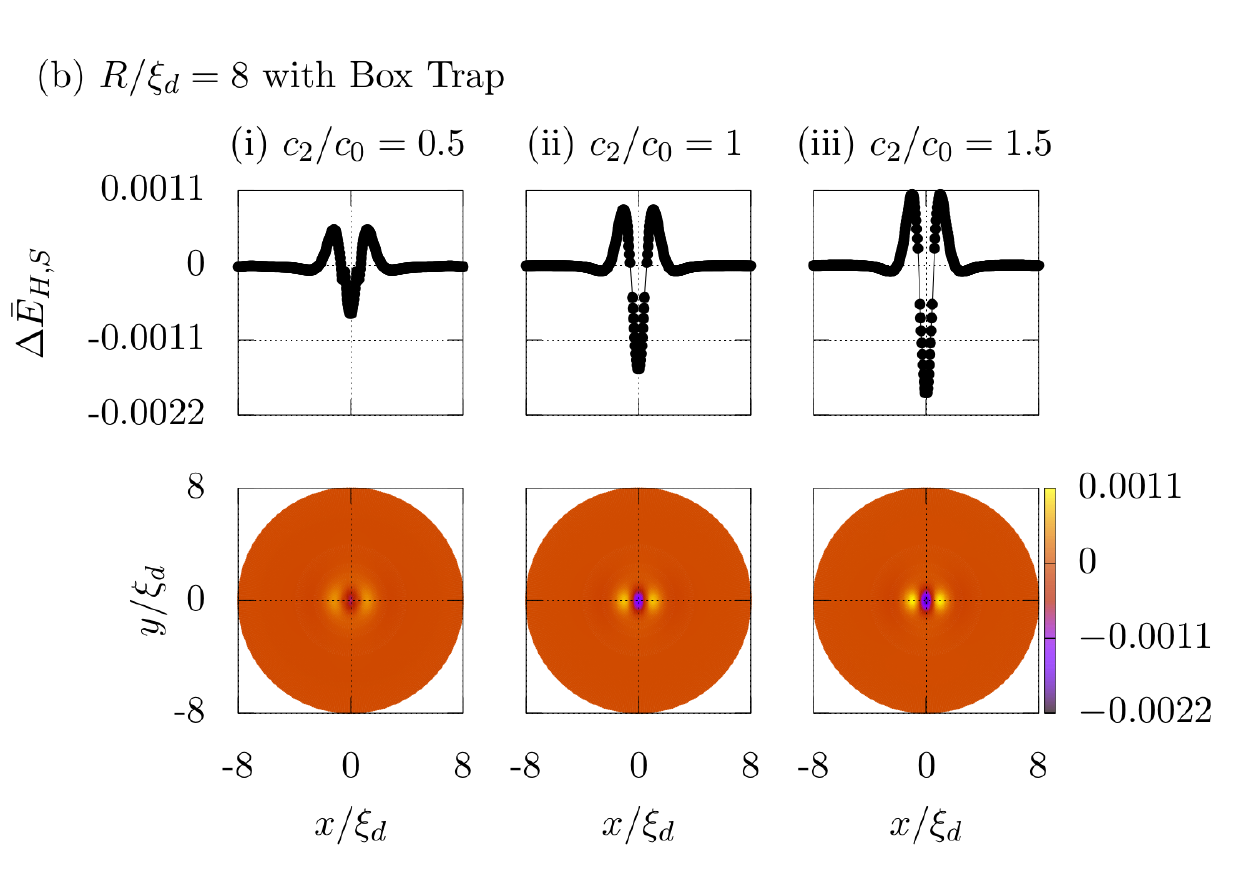}
\label{nt_R8_var_c2}
}
\caption{
\emph{$\Delta \bar{E}_{H, S}$ for various $c_2 / c_0$ with box trap.} 
The upper panel  shows $\Delta \bar{E}_{H, S}$ along the $x / \xi_d$ axis, and the lower panel 
displays 2D contour plots of $\Delta \bar{E}_{H, S}$.
The critical values $(c_2/c_0)_{\rm cr}$ are 1.9 and 1.2 for $R / \xi_d = 4$ and 8 with box trap, respectively.
As $c_2 / c_0$ decreases, $\Delta \bar{E}_{H, S}$ decreases due to the spin interaction energy.
}
\label{nt_R4_R8_en_den_fig}
\end{figure*}

To assess the dependence of the  energetical stability against vortex dissociation on $\left( R / \xi_d, c_2 / c_0 \right)$, we fix $ R / \xi_d = 4 $ and $c_0 > 0$, and obtained the data in Fig.~\ref{fig2} (a). 
One observes that $\Delta E_{H,S}$ has a critical value $(c_2/c_0)_{\rm cr} $ where the dissociation becomes unfavorable for $c_2 / c_0 > (c_2/c_0)_{\rm cr} $.
It has previously been established that the critical value $ (c_2 / c_0)_{\rm cr} = 1$ for an infinitely large system, implying that the dissociation becomes unfavorable for $ c_2 > c_0 $ (assuming as usual 
that $c_0>0$)~\cite{EtoInteraction}.  The latter reference performed an asymptotic expansion of the 
energy for $R \gg D \gg \xi$ where $\xi$ is a short distance cutoff~\footnote{We note that the couplings $g_{12},g_1,g_2$ defined in~\cite{EtoInteraction} are in our notation 
$g_{12}=c_0 - c_2$ and $g_1 = g_2 = c_0 + c_2$.}. 
This assumption clearly is not applicable when $R / \xi_d = 4$, which then leads to $(c_2/c_0)_{\rm cr} \neq 1$, cf.~Fig.~\ref{fig2}. However, even though the assumptions of~\cite{EtoInteraction} strictly speaking 
cannot be applied for our setup, the force between two oppositely charged HQVs is still repulsive for $R / \xi_d = 4$, because $E_H$ decreases as $D$ increases, consistent with the force description in~\cite{EtoInteraction} for $0 < c_2 / c_0 < 1$. 
Moreover, we predict that the intervortex force is still negative for $- 0.5 \le c_2 / c_0 < 0$. This clearly shows that the intervortex force formula of~\cite{EtoInteraction} does not hold for $c_2 < 0$ 
because according to the latter one would have attraction for $c_2 / c_0 < 0$ or $c_2 / c_0 > 1$. 
This discrepancy is due to the fact that~\cite{EtoInteraction} considered the $c_2 > 0$ region only,  where the AF phase represents the ground state. 
As $E_{\textrm{spin}}$ is proportional to $c_2$ where $E_{\textrm{spin}}$ is spin interaction energy, when $c_2 > 0$, the system minimizes $E_{\textrm{spin}}$ (equivalent to minimizing $\left\vert E_{\textrm{spin}} \right\vert$). However, when $c_2 < 0$, the wavefunction changes to maximize $\left\vert E_{\textrm{spin}} \right\vert$ while conserving $S_z$, which makes the ansatz of \cite{EtoInteraction} 
for the wavefunction invalid for $c_2 < 0$.


From Fig.~\ref{fig2}, as $R / \xi_d$ decreases, the critical value of $c_2 / c_0$ is expected to be shifted to larger $c_2 / c_0$ value. To check whether this expectation is true, we set $R / \xi_d = 8$ and changed $c_2$ while fixing $c_0$. 
Figure~\ref{fig2} (b) shows that $(c_2/c_0)_{\rm cr}$ value is about $1.2$, but it is larger than $(c_2/c_0)_{\rm cr}$ for $ R / \xi_d \rightarrow \infty $. Therefore, our assumption that $(c_2/c_0)_{\rm cr}$ is shifted to smaller value as $R / \xi_d$ increases is true for $4 \le R / \xi_d \le 8$. Together with the harmonic trap data (see section below), we summarize the critical $(c_2/c_0)_{\rm cr}$ values in table~\ref{table1}. 

To determine how $R / \xi_d$ affects $(c_2/c_0)_{\rm cr}$, as $(c_2/c_0)_{\rm cr}$ is determined by comparing $\tilde{E}_{H}$ and $\tilde{E}_{S}$, we correspondingly 
compared the (scaled) energy density for various $R / \xi_d$ and $c_2 / c_0$. Here, Eq.~\eqref{scaled_energy_density_def} defines the (scaled) energy density $\Delta \bar{E}_{H, S}$: 
\begin{eqnarray}
\Delta E_{H, S} && = \int_{0}^{2 \pi} d \varphi \int_{0}^{R_d} d r_d \; r_d \Delta \bar{E}_{H, S} .
\label{scaled_energy_density_def}
\end{eqnarray}
Here, $r_d \coloneqq r / \xi_d$ and $R_d \coloneqq R / \xi_d$. Note that $\Delta \bar{E}_{H, S}$ is dimensionless due to the definition of $\Delta E_{H, S}$, Eq.~\eqref{scaled_E_diff}.
Fig.~\ref{nt_R4_R8_en_den_fig} shows $\Delta \bar{E}_{H, S}$ for $R / \xi_d = 4$ and $8$ when $D / \xi_d = 1$.

Because $E_S$ and $E_{\textrm{ref}}$ do not depend on $c_2$ with a SQV in the AF phase
~\footnote{For SQVs, 
$Q_s = 0$ and Eq.~\eqref{meanprdiff_AF} becomes symmetric for $\tilde{A}_{\pm 1}$. Thus, $\tilde{A}_{1} = \tilde{A}_{-1}$, which renders the spin interaction energy part in Eq.~\eqref{energy_AF} zero. Hence, $E_s$ does not depend on $c_2$. With no vortex in the AF phase, $Q_n = Q_s = 0$ and it is easily  shown  that, according to Eqs.~\eqref{meanprdiff_AF} and~\eqref{energy_AF}, $E_{\textrm{ref}}$ does not depend on $c_2$.}, 
solely HQVs affect the shape of $\Delta \bar{E}_{H, S}$ when $c_2 / c_0$ changes. 
As one can see from Fig.~\ref{nt_R4_R8_en_den_fig}, the energy density  with HQVs is concentrated in their cores whereas it strongly decreases near the center when $c_2 / c_0$ increases.  
This is due to the combined effects of spin interaction energy, kinetic energy, and the phases of two oppositely charged HQVs. 
Since we assumed that two HQVs with $\left( q_n, q_s \right) = \left( 1/2, \pm 1/2 \right)$ are 
symmetrically placed at $\left( x, y \right) = \left( \pm \xi_d / 2,0 \right)$ ($D=\xi_d$), for $c_2 \rightarrow \infty$ keeping  $c_0$ fixed, from symmetry, normalization and total spin constraints, the following conclusions can be drawn:
(1) The spin healing length $\xi_s\propto 1/\sqrt{c_2}$ and the core size of a HQV decreases upon increasing $c_2$. 
Then, from~\cite{JiHQV}, in-between the cores of the two HQVs, the kinetic and spin interaction energy densities decrease to negligibly small values, and the density-density interaction contribution becomes constant. 
(2) While the spin interaction energy tends to minimize the difference of $\tilde{A}_{\pm 1}$ when $c_2 > 0$, due to phase constraints, $\tilde{A}^2_{1} - \tilde{A}^2_{-1}$ cannot be zero near the cores of HQVs. Hence, the spin interaction energy density in the cores of HQVs increases. 
Therefore, the double peaks in $\Delta \bar{E}_{H, S}$ become increasingly narrow when $c_2 / c_0$ increases, which leads to $\Delta E_{H, S} > 0$ for $c_2 / c_0 > \left( c_2 / c_0 \right)_{\textrm{cr}}$.
For smaller $R / \xi_d$, the cores of HQVs become relatively larger. As a result, $c_2 / c_0$ must increase to make the peaks in $\Delta \bar{E}_{H, S}$ more narrow. This constitutes the reason 
for $\left( c_2 / c_0 \right)_{\textrm{cr}}$ being increased when $R / \xi_d$ decreases.

\begin{figure}[b]
\centering
\includegraphics[width=0.3\textwidth]{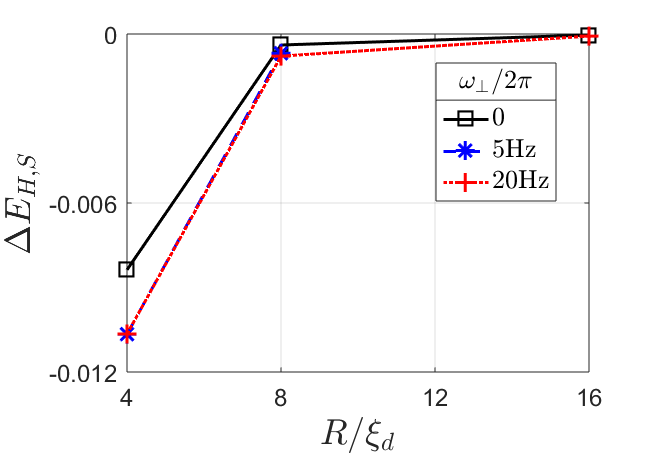}
\caption{\emph{Scaled energy difference $\Delta E_{H,S}$ as a function of $ R / \xi_d $ for $D / \xi_d = 1$, for box and harmonic traps.} 
Here, $c_2 = c_0$. Values of $R / \xi_d$ and $N$ are in Table~\ref{N_R_xi_d}.
We conclude that, relative to the box trap potential, the harmonic trap potential facilitates dissociation. 
}
\label{fig5}
\end{figure}

\begin{figure}[hbt]
\centering
\includegraphics[width=0.241\textwidth]{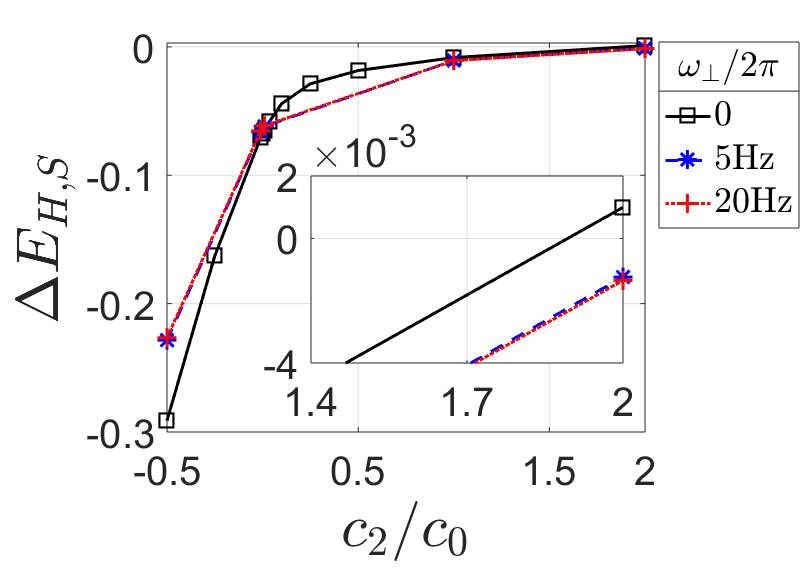}
\includegraphics[width=0.229\textwidth]{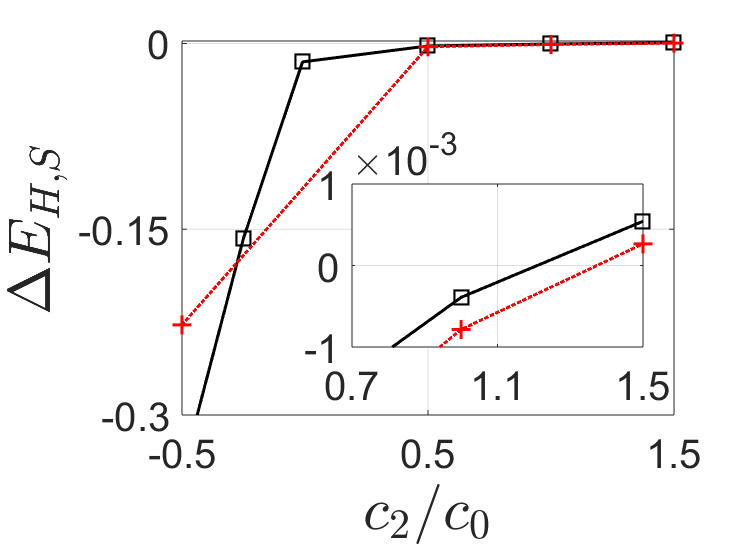}
\caption{\emph{Scaled energy difference $\Delta E_{H,S}$ as a function of $ c_2 / c_0 $} for (a) $R / \xi_d = 4$ and (b) $R / \xi_d = 8$ for box ($\omega_{\perp} = 0$) and harmonic trap.
Black dots are for the box trap, blue and red dots are harmonic traps with 
$\omega_{\perp} / 2 \pi = 5\, \textrm{Hz}$, and $\omega_{\perp} / 2 \pi = 20\, \textrm{Hz}$, 
respectively.
The energetical stability against vortex dissociation depends on $c_2 / c_0$ and the type of trap potential. 
The inset shows the critical crossing of $c_2 / c_0$;  
Horizontal and vertical axis are identical to the main plot}
\label{fig6}
\end{figure}

\begin{figure*}[hbt]
\centering
\subfigure{
\includegraphics[width=0.49\linewidth]{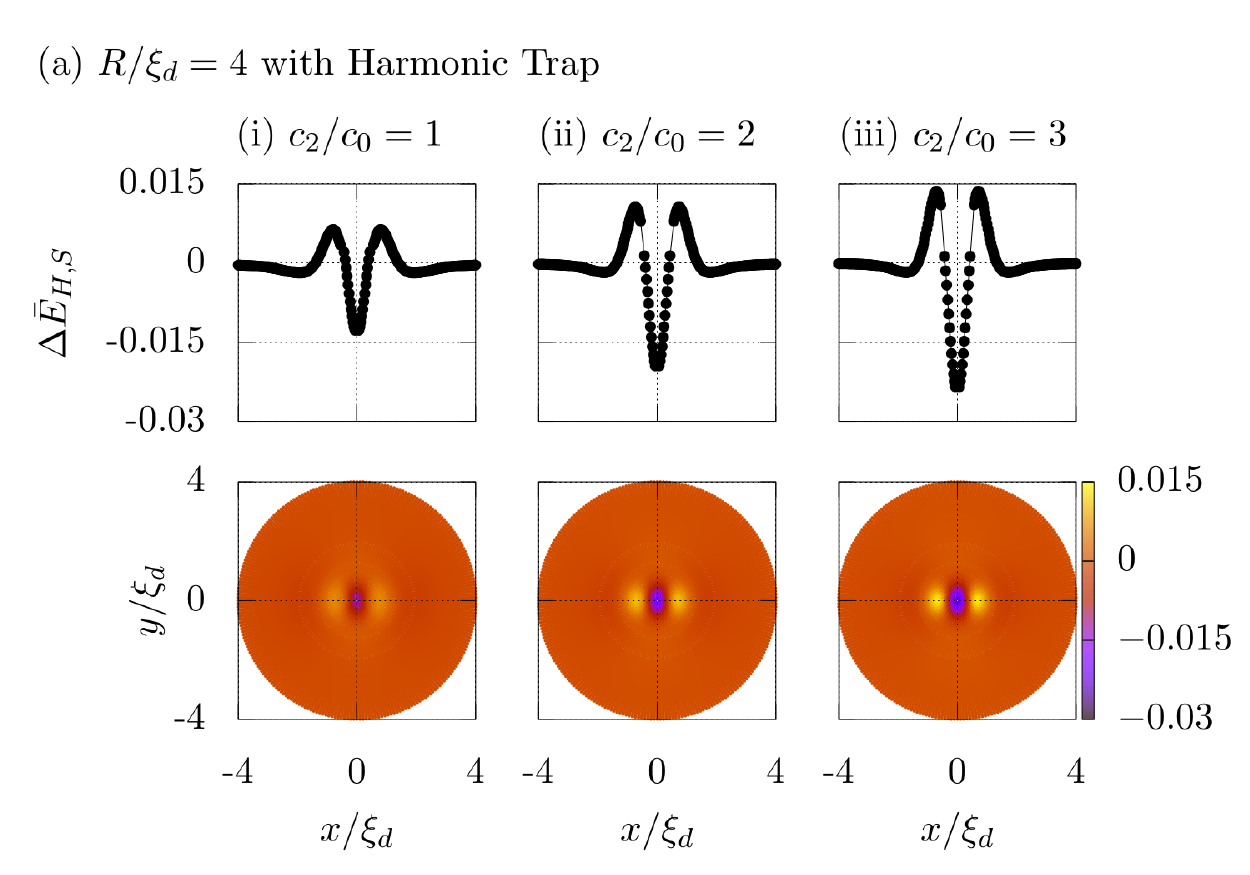}
\label{tr_R4_var_c2}
}
\hspace*{-1em}
\subfigure{
\includegraphics[width=0.49\linewidth]{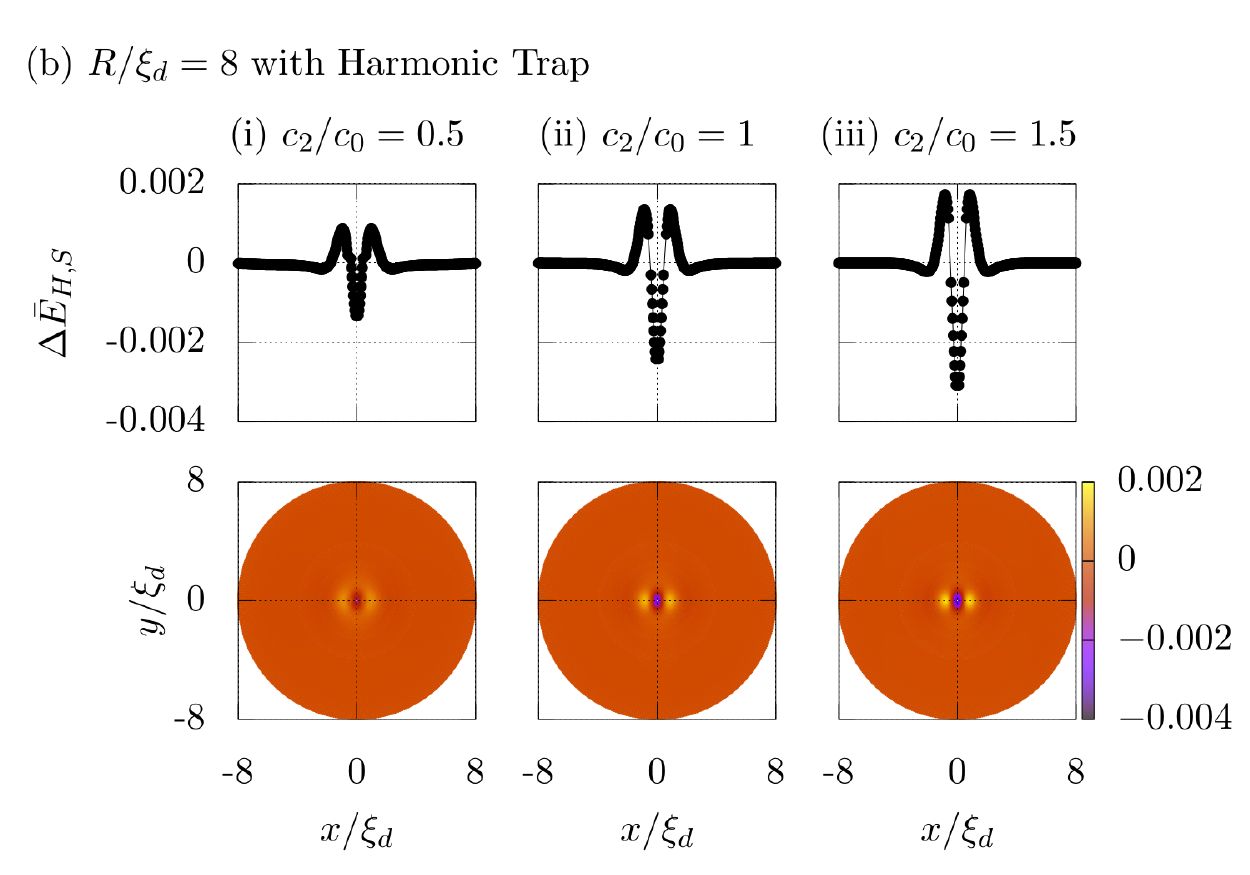}
\label{tr_R8_var_c2}
}
\caption{
\emph{$\Delta \bar{E}_{H, S}$ for various $c_2 / c_0$ with harmonic trap.} 
The upper panel  shows $\Delta \bar{E}_{H, S}$ along the $x / \xi_d$ axis, and the lower panel 
displays 2D contour plots of $\Delta \bar{E}_{H, S}$.
The critical value $(c_2/c_0)_{\rm cr}$ is 2.3 and 1.4 for $R / \xi_d = 4$ and 8, 
respectively, cf.~Fig.~\ref{nt_R4_R8_en_den_fig} for the box trap.
}
\label{tr_R4_R8_en_den_fig}
\end{figure*}

\subsection{\label{harmonic_data}Harmonic Trapping}

We now consider a harmonic trap potential in the $x$-$y$ plane. In a  quasi-2D spinor gas with $\omega_x = \omega_y = \omega_{\perp} > 0$, we have $\omega_{\perp} \ll \omega_z$.
Setting $\omega = \omega_{\perp}$, we have $\nu = 1$ in Eq.~\eqref{meanprdiff_AF} and~\eqref{energy_AF}. Also, $R$ is determined by the TF radius, Eq.~\eqref{TFRad2D_AF}. Note that, 
according to Eqs.~\eqref{xid_AF} and~\eqref{TFRad2D_AF}, $ R_{\textrm{TF}} / \xi_d = \left( R_{\textrm{TF}} / l_{\perp} \right)^2 / 2 = \left( R_{\textrm{TF}} / l \right)^2 / 2$ when $\nu = 1$. As $0 \le \tilde{r} \le \tilde{R} = R / l_{\perp}$ for $\nu = 1$, the (scaled) energy difference is independent of $\omega_{\perp}$. While we thus only have to solve Eq.~\eqref{meanprdiff_AF} for one specific value of $\omega_{\perp}$ in order to assess the energetical stability against vortex dissociation, to verify the validity of our numerical code, we solved Eq.~\eqref{meanprdiff_AF} for two different values of $\omega_{\perp}$.
Solving Eq.~\eqref{meanprdiff_AF} in a harmonic trap is more time-consuming when compared to box traps (at the same lateral size), due to the boundary condition imposed on the wavefunction.
For box traps, the boundary condition is simply $\psi = 0$ when $r = R$. However, for harmonic traps, there is no finite value of $R_c$ where $\psi = 0$ for $r = R_c$. The TF radius 
equals $R_c$ by definition only if we neglect the kinetic term in the GP equation.
Therefore, if one sets $\psi = 0$ at $r = R_{a, c}$ where $R_{a, c} \ge R_{\textrm{TF}}$ is some finite positive value in order to approximate the boundary when a harmonic trap is used with  a corresponding 
TF radius $R_{\textrm{TF}}$, near $R_{a, c}$ the calculation burden increases in order to 
achieve $\psi \simeq 0$ at $r = R_{a, c}$, even though $\psi$ does not change significantly for $r \ge R_{\textrm{TF}}$.
Hence, to reduce the calculation time and memory cost, we set $ R=R_{\textrm{TF}} + l $ and performed calculations for $R / \xi_d = 4, 8$.

Table~\ref{N_R_xi_d} contains the $R / \xi_d$, $N$, $\omega_{\perp}$ values which were investigated
for the harmonic trap.
First, fixing $c_2 = c_0$, we changed $N$ and obtained what is displayed in Fig.~\ref{fig5}, from which 
we conclude that vortex dissociation is energetically more favorable in harmonic than in box traps.  As discussed in the above, 
using Eqs.~\eqref{meanprdiff_AF},~\eqref{energy_AF}, and~\eqref{TFRad2D_AF}, 
$\Delta E_{H, S}$ is independent of the harmonic trap frequency $\omega_{\perp}$ once $\omega_{\perp} > 0$. To investigate whether $(c_2/c_0)_{\rm cr}$ depends on the type of trap potential, we fixed $R / \xi_d = 4$ and changed $c_2$ while fixing $c_0$ (see Fig~\ref{fig6}).
Again, Fig.~\ref{fig6} states that spin-spin coupling driven vortex dissociation is energetically more likely in harmonic traps when compared to box traps, the critical ratio $(c_2/c_0)_{\rm cr}$ being smaller in the 
box trap. 
By plotting $\Delta \bar{E}_{H, S}$ in a harmonic trap, Fig.~\ref{tr_R4_R8_en_den_fig}, we find that $\Delta \bar{E}_{H, S}$ has double peaks, as in the box trap, Fig.~\ref{nt_R4_R8_en_den_fig}. 

Taking into account the harmonic trap energy, the latter is for two HQVs smaller than with a SQV at the center; their difference however decreases when $c_2 \rightarrow \infty$. Hence, for the same $R / \xi_d$, the critical $(c_2/c_0)_{\rm cr} $ in harmonic traps is larger than in box traps, as table~\ref{table1} demonstrates. 

\begin{table}[h]
\caption{\label{table1}
The critical $(c_2/c_0)_{\rm cr}$ for box and harmonic traps.}
\begin{ruledtabular}
\begin{tabular}{cccc}
\multicolumn{2}{c}{Box trap} & \multicolumn{2}{c}{Harmonic trap} \\
$R / \xi_d$ & $(c_2/c_0)_{\rm cr}$ & $R / \xi_d$ & $(c_2/c_0)_{\rm cr}$ \\
\hline
4 & $ 1.9 $ & 4 & $ 2.3 $ \\
8 & $ 1.19 $ & 8 & $ 1.38 $ 
\end{tabular}
\end{ruledtabular}
\end{table}

\section{Conclusion}

Using mean-field theory, we numerically 
obtained the critical value of the ratio of spin-spin and density-density couplings, $(c_2/c_0)_{\rm cr}$, 
for SQV dissociation into two HQVs to take place in a trapped spin-one condensate. 
Vortex dissociation was demonstrated to be energetically disfavored when $ c_2 / c_0 > (c_2/c_0)_{\rm cr} $ given $c_0 > 0$ and the system is in the AF phase. With a box trap potential, and hence in a relatively homogeneous situation, we obtained consistency with the results of \cite{EtoInteraction}. Moreover, our results predict that the intervortex force between two oppositely charged HQVs is repulsive for $c_2 / c_0 < (c_2/c_0)_{\rm cr} $. This, in particular, generally implies that it remains repulsive for $c_2 / c_0 < 0$. 
Furthermore, vortex dissociation was shown to be energetically more favorable in a harmonic than in a box trap, in the sense that the critical $(c_2 / c_0)_{\rm cr}$ in harmonically trapped gases is {\em larger} than in box traps. 
For harmonic traps, we have also shown that $(c_2/c_0)_{\rm cr}$ does not depend on the trap frequency $\omega_{\perp}$ in the plane, and that $(c_2 / c_0)_{cr}$ {\em increases} as $R / \xi_d$ 
{\em decreases}, implying that stronger confinement necessitates larger spin-spin interaction to suppress the dissociation process in smaller systems.

While current experiments on spinor gases (for an overview see, e.g.,~\cite{Stamper-Kurn}) all 
operate at the untuned, atomically pre-given 
small values of $|c_2/c_0|$, both in sodium ($c_2>0$) as well as in rubidium ($c_2<0$), 
our predictions for the critical values of $c_2/c_0$ 
can be experimentally verified by tuning couplings using the microwave and radio-frequency techniques 
suggested in various theoretical proposals~\cite{changing_a0_by_microwave,a0_change_via_radio,Cheianov}.

Finally, our calculations provide 
a valuable benchmark for the accuracy of the numerically highly demanding solution of coupled spinor GP equations, and hence for their predictive power regarding the dynamics of 
topological defects in ultracold quantum gases. The latter is here represented by a sensitive prediction for  the critical dissociation point for single-quantum vortices into half-quantum vortices for mesoscopic samples. In such a process,  an intricate interplay of various (interaction and single-particle) terms in the energy functional becomes important, and mean-field theory is tested sensitively in the 
presence of many such competing terms. 

\bigskip
\begin{acknowledgments}
We thank Yong-il Shin for helpful discussions on his spinor condensate experiments.
We employed supercomputing resources of the Supercomputing Center of the Korea Institute of Science and Technology Information. This work was supported by the National Research Foundation of Korea Grant Nos.~NRF-2015-033908 (Global Ph.D.~Fellowship Program) and 2017R1A2A2A05001422
(Core Research Program). 
\end{acknowledgments}

\bigskip


\bibliography{cdct21}

\begin{thebibliography}{50}%
\makeatletter
\providecommand \@ifxundefined [1]{%
 \@ifx{#1\undefined}
}%
\providecommand \@ifnum [1]{%
 \ifnum #1\expandafter \@firstoftwo
 \else \expandafter \@secondoftwo
 \fi
}%
\providecommand \@ifx [1]{%
 \ifx #1\expandafter \@firstoftwo
 \else \expandafter \@secondoftwo
 \fi
}%
\providecommand \natexlab [1]{#1}%
\providecommand \enquote  [1]{``#1''}%
\providecommand \bibnamefont  [1]{#1}%
\providecommand \bibfnamefont [1]{#1}%
\providecommand \citenamefont [1]{#1}%
\providecommand \href@noop [0]{\@secondoftwo}%
\providecommand \href [0]{\begingroup \@sanitize@url \@href}%
\providecommand \@href[1]{\@@startlink{#1}\@@href}%
\providecommand \@@href[1]{\endgroup#1\@@endlink}%
\providecommand \@sanitize@url [0]{\catcode `\\12\catcode `\$12\catcode
  `\&12\catcode `\#12\catcode `\^12\catcode `\_12\catcode `\%12\relax}%
\providecommand \@@startlink[1]{}%
\providecommand \@@endlink[0]{}%
\providecommand \url  [0]{\begingroup\@sanitize@url \@url }%
\providecommand \@url [1]{\endgroup\@href {#1}{\urlprefix }}%
\providecommand \urlprefix  [0]{URL }%
\providecommand \Eprint [0]{\href }%
\providecommand \doibase [0]{http://dx.doi.org/}%
\providecommand \selectlanguage [0]{\@gobble}%
\providecommand \bibinfo  [0]{\@secondoftwo}%
\providecommand \bibfield  [0]{\@secondoftwo}%
\providecommand \translation [1]{[#1]}%
\providecommand \BibitemOpen [0]{}%
\providecommand \bibitemStop [0]{}%
\providecommand \bibitemNoStop [0]{.\EOS\space}%
\providecommand \EOS [0]{\spacefactor3000\relax}%
\providecommand \BibitemShut  [1]{\csname bibitem#1\endcsname}%
\let\auto@bib@innerbib\@empty
\bibitem [{\citenamefont {Kibble}(1976)}]{Kibble}%
  \BibitemOpen
  \bibfield  {author} {\bibinfo {author} {\bibfnamefont {T.~W.~B.}\
  \bibnamefont {Kibble}},\ }\bibfield  {title} {\enquote {\bibinfo {title}
  {{Topology of cosmic domains and strings}},}\ }\href
  {http://stacks.iop.org/0305-4470/9/i=8/a=029} {\bibfield  {journal} {\bibinfo
   {journal} {Journal of Physics A: Mathematical and General}\ }\textbf
  {\bibinfo {volume} {9}},\ \bibinfo {pages} {1387} (\bibinfo {year}
  {1976})}\BibitemShut {NoStop}%
\bibitem [{\citenamefont {Mermin}(1979)}]{Mermin}%
  \BibitemOpen
  \bibfield  {author} {\bibinfo {author} {\bibfnamefont {N.~D.}\ \bibnamefont
  {Mermin}},\ }\bibfield  {title} {\enquote {\bibinfo {title} {{The topological
  theory of defects in ordered media}},}\ }\href {\doibase
  10.1103/RevModPhys.51.591} {\bibfield  {journal} {\bibinfo  {journal} {Rev.
  Mod. Phys.}\ }\textbf {\bibinfo {volume} {51}},\ \bibinfo {pages} {591--648}
  (\bibinfo {year} {1979})}\BibitemShut {NoStop}%
\bibitem [{\citenamefont {{Volovik}}\ and\ \citenamefont
  {{Mineev}}(1976)}]{Grisha}%
  \BibitemOpen
  \bibfield  {author} {\bibinfo {author} {\bibfnamefont {G.~E.}\ \bibnamefont
  {{Volovik}}}\ and\ \bibinfo {author} {\bibfnamefont {V.~P.}\ \bibnamefont
  {{Mineev}}},\ }\bibfield  {title} {\enquote {\bibinfo {title} {{Line and
  point singularities in superfluid He$^{3}$}},}\ }\href@noop {} {\bibfield
  {journal} {\bibinfo  {journal} {Soviet Journal of Experimental and
  Theoretical Physics Letters}\ }\textbf {\bibinfo {volume} {24}},\ \bibinfo
  {pages} {561--563} (\bibinfo {year} {1976})}\BibitemShut {NoStop}%
\bibitem [{\citenamefont {Salomaa}\ and\ \citenamefont
  {Volovik}(1987)}]{Salomaa}%
  \BibitemOpen
  \bibfield  {author} {\bibinfo {author} {\bibfnamefont {M.~M.}\ \bibnamefont
  {Salomaa}}\ and\ \bibinfo {author} {\bibfnamefont {G.~E.}\ \bibnamefont
  {Volovik}},\ }\bibfield  {title} {\enquote {\bibinfo {title} {{Quantized
  vortices in superfluid $^{3}\mathrm{He}$}},}\ }\href {\doibase
  10.1103/RevModPhys.59.533} {\bibfield  {journal} {\bibinfo  {journal} {Rev.
  Mod. Phys.}\ }\textbf {\bibinfo {volume} {59}},\ \bibinfo {pages} {533--613}
  (\bibinfo {year} {1987})}\BibitemShut {NoStop}%
\bibitem [{\citenamefont {Autti}\ \emph {et~al.}(2016)\citenamefont {Autti},
  \citenamefont {Dmitriev}, \citenamefont {M\"akinen}, \citenamefont
  {Soldatov}, \citenamefont {Volovik}, \citenamefont {Yudin}, \citenamefont
  {Zavjalov},\ and\ \citenamefont {Eltsov}}]{Autti}%
  \BibitemOpen
  \bibfield  {author} {\bibinfo {author} {\bibfnamefont {S.}~\bibnamefont
  {Autti}}, \bibinfo {author} {\bibfnamefont {V.~V.}\ \bibnamefont {Dmitriev}},
  \bibinfo {author} {\bibfnamefont {J.~T.}\ \bibnamefont {M\"akinen}}, \bibinfo
  {author} {\bibfnamefont {A.~A.}\ \bibnamefont {Soldatov}}, \bibinfo {author}
  {\bibfnamefont {G.~E.}\ \bibnamefont {Volovik}}, \bibinfo {author}
  {\bibfnamefont {A.~N.}\ \bibnamefont {Yudin}}, \bibinfo {author}
  {\bibfnamefont {V.~V.}\ \bibnamefont {Zavjalov}}, \ and\ \bibinfo {author}
  {\bibfnamefont {V.~B.}\ \bibnamefont {Eltsov}},\ }\bibfield  {title}
  {\enquote {\bibinfo {title} {{Observation of Half-Quantum Vortices in
  Topological Superfluid $^{3}\mathrm{He}$}},}\ }\href {\doibase
  10.1103/PhysRevLett.117.255301} {\bibfield  {journal} {\bibinfo  {journal}
  {Phys. Rev. Lett.}\ }\textbf {\bibinfo {volume} {117}},\ \bibinfo {pages}
  {255301} (\bibinfo {year} {2016})}\BibitemShut {NoStop}%
\bibitem [{\citenamefont {Ivanov}(2001)}]{Ivanov}%
  \BibitemOpen
  \bibfield  {author} {\bibinfo {author} {\bibfnamefont {D.~A.}\ \bibnamefont
  {Ivanov}},\ }\bibfield  {title} {\enquote {\bibinfo {title} {{Non-Abelian
  Statistics of Half-Quantum Vortices in $\mathit{p}$-Wave Superconductors}},}\
  }\href {\doibase 10.1103/PhysRevLett.86.268} {\bibfield  {journal} {\bibinfo
  {journal} {Phys. Rev. Lett.}\ }\textbf {\bibinfo {volume} {86}},\ \bibinfo
  {pages} {268--271} (\bibinfo {year} {2001})}\BibitemShut {NoStop}%
\bibitem [{\citenamefont {Jang}\ \emph {et~al.}(2011)\citenamefont {Jang},
  \citenamefont {Ferguson}, \citenamefont {Vakaryuk}, \citenamefont {Budakian},
  \citenamefont {Chung}, \citenamefont {Goldbart},\ and\ \citenamefont
  {Maeno}}]{Jang}%
  \BibitemOpen
  \bibfield  {author} {\bibinfo {author} {\bibfnamefont {J.}~\bibnamefont
  {Jang}}, \bibinfo {author} {\bibfnamefont {D.~G.}\ \bibnamefont {Ferguson}},
  \bibinfo {author} {\bibfnamefont {V.}~\bibnamefont {Vakaryuk}}, \bibinfo
  {author} {\bibfnamefont {R.}~\bibnamefont {Budakian}}, \bibinfo {author}
  {\bibfnamefont {S.~B.}\ \bibnamefont {Chung}}, \bibinfo {author}
  {\bibfnamefont {P.~M.}\ \bibnamefont {Goldbart}}, \ and\ \bibinfo {author}
  {\bibfnamefont {Y.}~\bibnamefont {Maeno}},\ }\bibfield  {title} {\enquote
  {\bibinfo {title} {{Observation of Half-Height Magnetization Steps in
  Sr$_2$RuO$_4$}},}\ }\href {\doibase 10.1126/science.1193839} {\bibfield
  {journal} {\bibinfo  {journal} {Science}\ }\textbf {\bibinfo {volume}
  {331}},\ \bibinfo {pages} {186--188} (\bibinfo {year} {2011})}\BibitemShut
  {NoStop}%
\bibitem [{\citenamefont {Lagoudakis}\ \emph {et~al.}(2009)\citenamefont
  {Lagoudakis}, \citenamefont {Ostatnick{\'y}}, \citenamefont {Kavokin},
  \citenamefont {Rubo}, \citenamefont {Andr{\'e}},\ and\ \citenamefont
  {Deveaud-Pl{\'e}dran}}]{HQV_observation_1st}%
  \BibitemOpen
  \bibfield  {author} {\bibinfo {author} {\bibfnamefont {K.~G.}\ \bibnamefont
  {Lagoudakis}}, \bibinfo {author} {\bibfnamefont {T.}~\bibnamefont
  {Ostatnick{\'y}}}, \bibinfo {author} {\bibfnamefont {A.~V.}\ \bibnamefont
  {Kavokin}}, \bibinfo {author} {\bibfnamefont {Y.~G.}\ \bibnamefont {Rubo}},
  \bibinfo {author} {\bibfnamefont {R.}~\bibnamefont {Andr{\'e}}}, \ and\
  \bibinfo {author} {\bibfnamefont {B.}~\bibnamefont {Deveaud-Pl{\'e}dran}},\
  }\bibfield  {title} {\enquote {\bibinfo {title} {{Observation of Half-Quantum
  Vortices in an Exciton-Polariton Condensate}},}\ }\href {\doibase
  10.1126/science.1177980} {\bibfield  {journal} {\bibinfo  {journal}
  {Science}\ }\textbf {\bibinfo {volume} {326}},\ \bibinfo {pages} {974--976}
  (\bibinfo {year} {2009})}\BibitemShut {NoStop}%
\bibitem [{\citenamefont {Manni}\ \emph {et~al.}(2012)\citenamefont {Manni},
  \citenamefont {Lagoudakis}, \citenamefont {Liew}, \citenamefont {Andr{\'e}},
  \citenamefont {Savona},\ and\ \citenamefont {Deveaud}}]{SQV-HQV_dissoc_exp}%
  \BibitemOpen
  \bibfield  {author} {\bibinfo {author} {\bibfnamefont {F.}~\bibnamefont
  {Manni}}, \bibinfo {author} {\bibfnamefont {K.~G.}\ \bibnamefont
  {Lagoudakis}}, \bibinfo {author} {\bibfnamefont {T.~C.~H.}\ \bibnamefont
  {Liew}}, \bibinfo {author} {\bibfnamefont {R.}~\bibnamefont {Andr{\'e}}},
  \bibinfo {author} {\bibfnamefont {V.}~\bibnamefont {Savona}}, \ and\ \bibinfo
  {author} {\bibfnamefont {B.}~\bibnamefont {Deveaud}},\ }\bibfield  {title}
  {\enquote {\bibinfo {title} {{{Dissociation dynamics of singly charged
  vortices into half-quantum vortex pairs}}},}\ }\href {\doibase
  10.1038/ncomms2310} {\bibfield  {journal} {\bibinfo  {journal} {Nature
  Communications}\ }\textbf {\bibinfo {volume} {3}},\ \bibinfo {pages} {1309}
  (\bibinfo {year} {2012})}\BibitemShut {NoStop}%
\bibitem [{\citenamefont {Leonhardt}\ and\ \citenamefont
  {Volovik}(2000)}]{Leonhardt}%
  \BibitemOpen
  \bibfield  {author} {\bibinfo {author} {\bibfnamefont {U.}~\bibnamefont
  {Leonhardt}}\ and\ \bibinfo {author} {\bibfnamefont {G.~E.}\ \bibnamefont
  {Volovik}},\ }\bibfield  {title} {\enquote {\bibinfo {title} {{How to create
  an Alice string (half-quantum vortex) in a vector Bose-Einstein
  condensate}},}\ }\href {\doibase 10.1134/1.1312008} {\bibfield  {journal}
  {\bibinfo  {journal} {Journal of Experimental and Theoretical Physics
  Letters}\ }\textbf {\bibinfo {volume} {72}},\ \bibinfo {pages} {46--48}
  (\bibinfo {year} {2000})}\BibitemShut {NoStop}%
\bibitem [{\citenamefont {Ruostekoski}\ and\ \citenamefont
  {Anglin}(2003)}]{monopole_core_HQV_ring}%
  \BibitemOpen
  \bibfield  {author} {\bibinfo {author} {\bibfnamefont {J.}~\bibnamefont
  {Ruostekoski}}\ and\ \bibinfo {author} {\bibfnamefont {J.~R.}\ \bibnamefont
  {Anglin}},\ }\bibfield  {title} {\enquote {\bibinfo {title} {{Monopole Core
  Instability and Alice Rings in Spinor Bose-Einstein Condensates}},}\ }\href
  {\doibase 10.1103/PhysRevLett.91.190402} {\bibfield  {journal} {\bibinfo
  {journal} {Phys. Rev. Lett.}\ }\textbf {\bibinfo {volume} {91}},\ \bibinfo
  {pages} {190402} (\bibinfo {year} {2003})}\BibitemShut {NoStop}%
\bibitem [{\citenamefont {Matthews}\ \emph {et~al.}(1999)\citenamefont
  {Matthews}, \citenamefont {Anderson}, \citenamefont {Haljan}, \citenamefont
  {Hall}, \citenamefont {Wieman},\ and\ \citenamefont
  {Cornell}}]{1st_vortex_creation}%
  \BibitemOpen
  \bibfield  {author} {\bibinfo {author} {\bibfnamefont {M.~R.}\ \bibnamefont
  {Matthews}}, \bibinfo {author} {\bibfnamefont {B.~P.}\ \bibnamefont
  {Anderson}}, \bibinfo {author} {\bibfnamefont {P.~C.}\ \bibnamefont
  {Haljan}}, \bibinfo {author} {\bibfnamefont {D.~S.}\ \bibnamefont {Hall}},
  \bibinfo {author} {\bibfnamefont {C.~E.}\ \bibnamefont {Wieman}}, \ and\
  \bibinfo {author} {\bibfnamefont {E.~A.}\ \bibnamefont {Cornell}},\
  }\bibfield  {title} {\enquote {\bibinfo {title} {{Vortices in a Bose-Einstein
  Condensate}},}\ }\href {\doibase 10.1103/PhysRevLett.83.2498} {\bibfield
  {journal} {\bibinfo  {journal} {Phys. Rev. Lett.}\ }\textbf {\bibinfo
  {volume} {83}},\ \bibinfo {pages} {2498--2501} (\bibinfo {year}
  {1999})}\BibitemShut {NoStop}%
\bibitem [{\citenamefont {Madison}\ \emph {et~al.}(2000)\citenamefont
  {Madison}, \citenamefont {Chevy}, \citenamefont {Wohlleben},\ and\
  \citenamefont {Dalibard}}]{Madison}%
  \BibitemOpen
  \bibfield  {author} {\bibinfo {author} {\bibfnamefont {K.~W.}\ \bibnamefont
  {Madison}}, \bibinfo {author} {\bibfnamefont {F.}~\bibnamefont {Chevy}},
  \bibinfo {author} {\bibfnamefont {W.}~\bibnamefont {Wohlleben}}, \ and\
  \bibinfo {author} {\bibfnamefont {J.}~\bibnamefont {Dalibard}},\ }\bibfield
  {title} {\enquote {\bibinfo {title} {{Vortex Formation in a Stirred
  Bose-Einstein Condensate}},}\ }\href {\doibase 10.1103/PhysRevLett.84.806}
  {\bibfield  {journal} {\bibinfo  {journal} {Phys. Rev. Lett.}\ }\textbf
  {\bibinfo {volume} {84}},\ \bibinfo {pages} {806--809} (\bibinfo {year}
  {2000})}\BibitemShut {NoStop}%
\bibitem [{\citenamefont {Kasamatsu}\ \emph {et~al.}(2005)\citenamefont
  {Kasamatsu}, \citenamefont {Tsubota},\ and\ \citenamefont
  {Ueda}}]{vortices_in_BEC}%
  \BibitemOpen
  \bibfield  {author} {\bibinfo {author} {\bibfnamefont {K.}~\bibnamefont
  {Kasamatsu}}, \bibinfo {author} {\bibfnamefont {M.}~\bibnamefont {Tsubota}},
  \ and\ \bibinfo {author} {\bibfnamefont {M.}~\bibnamefont {Ueda}},\
  }\bibfield  {title} {\enquote {\bibinfo {title} {{Vortices in Multicomponent
  Bose-Einstein condensates}},}\ }\href {\doibase 10.1142/S0217979205029602}
  {\bibfield  {journal} {\bibinfo  {journal} {International Journal of Modern
  Physics B}\ }\textbf {\bibinfo {volume} {19}},\ \bibinfo {pages} {1835}
  (\bibinfo {year} {2005})}\BibitemShut {NoStop}%
\bibitem [{\citenamefont {Kawaguchi}\ and\ \citenamefont
  {Ueda}(2012)}]{Uedareview}%
  \BibitemOpen
  \bibfield  {author} {\bibinfo {author} {\bibfnamefont {Yuki}\ \bibnamefont
  {Kawaguchi}}\ and\ \bibinfo {author} {\bibfnamefont {Masahito}\ \bibnamefont
  {Ueda}},\ }\bibfield  {title} {\enquote {\bibinfo {title} {{Spinor
  Bose--Einstein condensates}},}\ }\href {\doibase
  10.1016/j.physrep.2012.07.005} {\bibfield  {journal} {\bibinfo  {journal}
  {Physics Reports}\ }\textbf {\bibinfo {volume} {520}},\ \bibinfo {pages}
  {253--381} (\bibinfo {year} {2012})}\BibitemShut {NoStop}%
\bibitem [{\citenamefont {Ueda}(2014)}]{UedaTopo}%
  \BibitemOpen
  \bibfield  {author} {\bibinfo {author} {\bibfnamefont {Masahito}\
  \bibnamefont {Ueda}},\ }\bibfield  {title} {\enquote {\bibinfo {title}
  {{Topological aspects in spinor Bose--Einstein condensates}},}\ }\href
  {http://stacks.iop.org/0034-4885/77/i=12/a=122401} {\bibfield  {journal}
  {\bibinfo  {journal} {Reports on Progress in Physics}\ }\textbf {\bibinfo
  {volume} {77}},\ \bibinfo {pages} {122401} (\bibinfo {year}
  {2014})}\BibitemShut {NoStop}%
\bibitem [{\citenamefont {Ho}(1998)}]{Jason}%
  \BibitemOpen
  \bibfield  {author} {\bibinfo {author} {\bibfnamefont {Tin-Lun}\ \bibnamefont
  {Ho}},\ }\bibfield  {title} {\enquote {\bibinfo {title} {{Spinor Bose
  Condensates in Optical Traps}},}\ }\href {\doibase
  10.1103/PhysRevLett.81.742} {\bibfield  {journal} {\bibinfo  {journal} {Phys.
  Rev. Lett.}\ }\textbf {\bibinfo {volume} {81}},\ \bibinfo {pages} {742--745}
  (\bibinfo {year} {1998})}\BibitemShut {NoStop}%
\bibitem [{\citenamefont {Ohmi}\ and\ \citenamefont {Machida}(1998)}]{Ohmi}%
  \BibitemOpen
  \bibfield  {author} {\bibinfo {author} {\bibfnamefont {Tetsuo}\ \bibnamefont
  {Ohmi}}\ and\ \bibinfo {author} {\bibfnamefont {Kazushige}\ \bibnamefont
  {Machida}},\ }\bibfield  {title} {\enquote {\bibinfo {title} {{Bose-Einstein
  Condensation with Internal Degrees of Freedom in Alkali Atom Gases}},}\
  }\href {\doibase 10.1143/JPSJ.67.1822} {\bibfield  {journal} {\bibinfo
  {journal} {Journal of the Physical Society of Japan}\ }\textbf {\bibinfo
  {volume} {67}},\ \bibinfo {pages} {1822--1825} (\bibinfo {year}
  {1998})}\BibitemShut {NoStop}%
\bibitem [{\citenamefont {Zhou}(2001)}]{Zhou}%
  \BibitemOpen
  \bibfield  {author} {\bibinfo {author} {\bibfnamefont {Fei}\ \bibnamefont
  {Zhou}},\ }\bibfield  {title} {\enquote {\bibinfo {title} {{Spin Correlation
  and Discrete Symmetry in Spinor Bose-Einstein Condensates}},}\ }\href
  {\doibase 10.1103/PhysRevLett.87.080401} {\bibfield  {journal} {\bibinfo
  {journal} {Phys. Rev. Lett.}\ }\textbf {\bibinfo {volume} {87}},\ \bibinfo
  {pages} {080401} (\bibinfo {year} {2001})}\BibitemShut {NoStop}%
\bibitem [{\citenamefont {Ji}\ \emph {et~al.}(2008)\citenamefont {Ji},
  \citenamefont {Liu}, \citenamefont {Song},\ and\ \citenamefont
  {Zhou}}]{JiHQV}%
  \BibitemOpen
  \bibfield  {author} {\bibinfo {author} {\bibfnamefont {An-Chun}\ \bibnamefont
  {Ji}}, \bibinfo {author} {\bibfnamefont {W.~M.}\ \bibnamefont {Liu}},
  \bibinfo {author} {\bibfnamefont {Jun~Liang}\ \bibnamefont {Song}}, \ and\
  \bibinfo {author} {\bibfnamefont {Fei}\ \bibnamefont {Zhou}},\ }\bibfield
  {title} {\enquote {\bibinfo {title} {{Dynamical Creation of Fractionalized
  Vortices and Vortex Lattices}},}\ }\href {\doibase
  10.1103/PhysRevLett.101.010402} {\bibfield  {journal} {\bibinfo  {journal}
  {Phys. Rev. Lett.}\ }\textbf {\bibinfo {volume} {101}},\ \bibinfo {pages}
  {010402} (\bibinfo {year} {2008})}\BibitemShut {NoStop}%
\bibitem [{\citenamefont {Fetter}(2014)}]{Fetter}%
  \BibitemOpen
  \bibfield  {author} {\bibinfo {author} {\bibfnamefont {Alexander~L.}\
  \bibnamefont {Fetter}},\ }\bibfield  {title} {\enquote {\bibinfo {title}
  {{Vortex dynamics in spin-orbit-coupled Bose-Einstein condensates}},}\ }\href
  {\doibase 10.1103/PhysRevA.89.023629} {\bibfield  {journal} {\bibinfo
  {journal} {Phys. Rev. A}\ }\textbf {\bibinfo {volume} {89}},\ \bibinfo
  {pages} {023629} (\bibinfo {year} {2014})}\BibitemShut {NoStop}%
\bibitem [{\citenamefont {Shirley}\ \emph {et~al.}(2014)\citenamefont
  {Shirley}, \citenamefont {Anderson}, \citenamefont {Clark},\ and\
  \citenamefont {Wilson}}]{Shirley}%
  \BibitemOpen
  \bibfield  {author} {\bibinfo {author} {\bibfnamefont {Wilbur~E.}\
  \bibnamefont {Shirley}}, \bibinfo {author} {\bibfnamefont {Brandon~M.}\
  \bibnamefont {Anderson}}, \bibinfo {author} {\bibfnamefont {Charles~W.}\
  \bibnamefont {Clark}}, \ and\ \bibinfo {author} {\bibfnamefont {Ryan~M.}\
  \bibnamefont {Wilson}},\ }\bibfield  {title} {\enquote {\bibinfo {title}
  {{Half-Quantum Vortex Molecules in a Binary Dipolar Bose Gas}},}\ }\href
  {\doibase 10.1103/PhysRevLett.113.165301} {\bibfield  {journal} {\bibinfo
  {journal} {Phys. Rev. Lett.}\ }\textbf {\bibinfo {volume} {113}},\ \bibinfo
  {pages} {165301} (\bibinfo {year} {2014})}\BibitemShut {NoStop}%
\bibitem [{\citenamefont {Symes}\ and\ \citenamefont {Blakie}(2017)}]{Symes}%
  \BibitemOpen
  \bibfield  {author} {\bibinfo {author} {\bibfnamefont {L.~M.}\ \bibnamefont
  {Symes}}\ and\ \bibinfo {author} {\bibfnamefont {P.~B.}\ \bibnamefont
  {Blakie}},\ }\bibfield  {title} {\enquote {\bibinfo {title} {Nematic ordering
  dynamics of an antiferromagnetic spin-1 condensate},}\ }\href {\doibase
  10.1103/PhysRevA.96.013602} {\bibfield  {journal} {\bibinfo  {journal} {Phys.
  Rev. A}\ }\textbf {\bibinfo {volume} {96}},\ \bibinfo {pages} {013602}
  (\bibinfo {year} {2017})}\BibitemShut {NoStop}%
\bibitem [{\citenamefont {Schwarz}(1982)}]{Schwarz}%
  \BibitemOpen
  \bibfield  {author} {\bibinfo {author} {\bibfnamefont {A.~S.}\ \bibnamefont
  {Schwarz}},\ }\bibfield  {title} {\enquote {\bibinfo {title} {{Field theories
  with no local conservation of the electric charge}},}\ }\href {\doibase
  https://doi.org/10.1016/0550-3213(82)90190-0} {\bibfield  {journal} {\bibinfo
   {journal} {Nuclear Physics B}\ }\textbf {\bibinfo {volume} {208}},\ \bibinfo
  {pages} {141--158} (\bibinfo {year} {1982})}\BibitemShut {NoStop}%
\bibitem [{\citenamefont {Hindmarsh}\ and\ \citenamefont
  {Kibble}(1995)}]{Hindmarsh}%
  \BibitemOpen
  \bibfield  {author} {\bibinfo {author} {\bibfnamefont {M.~B.}\ \bibnamefont
  {Hindmarsh}}\ and\ \bibinfo {author} {\bibfnamefont {T.~W.~B.}\ \bibnamefont
  {Kibble}},\ }\bibfield  {title} {\enquote {\bibinfo {title} {{Cosmic
  strings}},}\ }\href {http://stacks.iop.org/0034-4885/58/i=5/a=001} {\bibfield
   {journal} {\bibinfo  {journal} {Reports on Progress in Physics}\ }\textbf
  {\bibinfo {volume} {58}},\ \bibinfo {pages} {477} (\bibinfo {year}
  {1995})}\BibitemShut {NoStop}%
\bibitem [{\citenamefont {Volovik}(2009)}]{GrishaBook}%
  \BibitemOpen
  \bibfield  {author} {\bibinfo {author} {\bibfnamefont {G.~E.}\ \bibnamefont
  {Volovik}},\ }\href {https://books.google.de/books?id=6uj76kFJOHEC} {\emph
  {\bibinfo {title} {{The Universe in a Helium Droplet}}}},\ International
  Series of Monographs on Physics\ (\bibinfo  {publisher} {OUP Oxford},\
  \bibinfo {year} {2009})\BibitemShut {NoStop}%
\bibitem [{\citenamefont {{Eto}}\ and\ \citenamefont
  {{Nitta}}(2018)}]{HQV-Quark}%
  \BibitemOpen
  \bibfield  {author} {\bibinfo {author} {\bibfnamefont {Minoru}\ \bibnamefont
  {{Eto}}}\ and\ \bibinfo {author} {\bibfnamefont {Muneto}\ \bibnamefont
  {{Nitta}}},\ }\bibfield  {title} {\enquote {\bibinfo {title} {{{Confinement
  of half-quantized vortices in coherently coupled Bose-Einstein condensates:
  Simulating quark confinement in a QCD-like theory}}},}\ }\href {\doibase
  10.1103/PhysRevA.97.023613} {\bibfield  {journal} {\bibinfo  {journal} {Phys.
  Rev. A}\ }\textbf {\bibinfo {volume} {97}},\ \bibinfo {pages} {023613}
  (\bibinfo {year} {2018})}\BibitemShut {NoStop}%
\bibitem [{\citenamefont {Tylutki}\ \emph {et~al.}(2016)\citenamefont
  {Tylutki}, \citenamefont {Pitaevski\v\i}, \citenamefont {Recati},\ and\
  \citenamefont {Stringari}}]{vortex_pair_confine_Rabi}%
  \BibitemOpen
  \bibfield  {author} {\bibinfo {author} {\bibfnamefont {Marek}\ \bibnamefont
  {Tylutki}}, \bibinfo {author} {\bibfnamefont {Lev~P.}\ \bibnamefont
  {Pitaevski\v\i}}, \bibinfo {author} {\bibfnamefont {Alessio}\ \bibnamefont
  {Recati}}, \ and\ \bibinfo {author} {\bibfnamefont {Sandro}\ \bibnamefont
  {Stringari}},\ }\bibfield  {title} {\enquote {\bibinfo {title} {{Confinement
  and precession of vortex pairs in coherently coupled Bose-Einstein
  condensates}},}\ }\href {\doibase 10.1103/PhysRevA.93.043623} {\bibfield
  {journal} {\bibinfo  {journal} {Phys. Rev. A}\ }\textbf {\bibinfo {volume}
  {93}},\ \bibinfo {pages} {043623} (\bibinfo {year} {2016})}\BibitemShut
  {NoStop}%
\bibitem [{\citenamefont {Seo}\ \emph {et~al.}(2015)\citenamefont {Seo},
  \citenamefont {Kang}, \citenamefont {Kwon},\ and\ \citenamefont
  {Shin}}]{ShinHQVExp}%
  \BibitemOpen
  \bibfield  {author} {\bibinfo {author} {\bibfnamefont {Sang~Won}\
  \bibnamefont {Seo}}, \bibinfo {author} {\bibfnamefont {Seji}\ \bibnamefont
  {Kang}}, \bibinfo {author} {\bibfnamefont {Woo~Jin}\ \bibnamefont {Kwon}}, \
  and\ \bibinfo {author} {\bibfnamefont {Yong-il}\ \bibnamefont {Shin}},\
  }\bibfield  {title} {\enquote {\bibinfo {title} {{Half-Quantum Vortices in an
  Antiferromagnetic Spinor Bose-Einstein Condensate}},}\ }\href {\doibase
  10.1103/PhysRevLett.115.015301} {\bibfield  {journal} {\bibinfo  {journal}
  {Phys. Rev. Lett.}\ }\textbf {\bibinfo {volume} {115}},\ \bibinfo {pages}
  {015301} (\bibinfo {year} {2015})}\BibitemShut {NoStop}%
\bibitem [{\citenamefont {Seo}\ \emph {et~al.}(2016)\citenamefont {Seo},
  \citenamefont {Kwon}, \citenamefont {Kang},\ and\ \citenamefont
  {Shin}}]{Collisional}%
  \BibitemOpen
  \bibfield  {author} {\bibinfo {author} {\bibfnamefont {Sang~Won}\
  \bibnamefont {Seo}}, \bibinfo {author} {\bibfnamefont {Woo~Jin}\ \bibnamefont
  {Kwon}}, \bibinfo {author} {\bibfnamefont {Seji}\ \bibnamefont {Kang}}, \
  and\ \bibinfo {author} {\bibfnamefont {Y.}~\bibnamefont {Shin}},\ }\bibfield
  {title} {\enquote {\bibinfo {title} {{Collisional Dynamics of Half-Quantum
  Vortices in a Spinor Bose-Einstein Condensate}},}\ }\href {\doibase
  10.1103/PhysRevLett.116.185301} {\bibfield  {journal} {\bibinfo  {journal}
  {Phys. Rev. Lett.}\ }\textbf {\bibinfo {volume} {116}},\ \bibinfo {pages}
  {185301} (\bibinfo {year} {2016})}\BibitemShut {NoStop}%
\bibitem [{\citenamefont {Choi}\ \emph {et~al.}(2012)\citenamefont {Choi},
  \citenamefont {Kwon},\ and\ \citenamefont {Shin}}]{Choi}%
  \BibitemOpen
  \bibfield  {author} {\bibinfo {author} {\bibfnamefont {Jae-yoon}\
  \bibnamefont {Choi}}, \bibinfo {author} {\bibfnamefont {Woo~Jin}\
  \bibnamefont {Kwon}}, \ and\ \bibinfo {author} {\bibfnamefont {Yong-il}\
  \bibnamefont {Shin}},\ }\bibfield  {title} {\enquote {\bibinfo {title}
  {{Observation of Topologically Stable 2D Skyrmions in an Antiferromagnetic
  Spinor Bose-Einstein Condensate}},}\ }\href {\doibase
  10.1103/PhysRevLett.108.035301} {\bibfield  {journal} {\bibinfo  {journal}
  {Phys. Rev. Lett.}\ }\textbf {\bibinfo {volume} {108}},\ \bibinfo {pages}
  {035301} (\bibinfo {year} {2012})}\BibitemShut {NoStop}%
\bibitem [{Not()}]{Note}%
  \BibitemOpen
  \href@noop {} {\ }\bibinfo {note} {\!\!We note that a similar dissociation
  process has been observed in polariton condensates \cite{SQV-HQV_dissoc_exp},
  where the pumped, nonequilibrium character of the system 
  modelization however however requires {\em driven-dissipative} spinor GP
  equations with some model assumptions.}\BibitemShut {Stop}%
\bibitem [{\citenamefont {Eto}\ \emph {et~al.}(2011)\citenamefont {Eto},
  \citenamefont {Kasamatsu}, \citenamefont {Nitta}, \citenamefont {Takeuchi},\
  and\ \citenamefont {Tsubota}}]{EtoInteraction}%
  \BibitemOpen
  \bibfield  {author} {\bibinfo {author} {\bibfnamefont {Minoru}\ \bibnamefont
  {Eto}}, \bibinfo {author} {\bibfnamefont {Kenichi}\ \bibnamefont
  {Kasamatsu}}, \bibinfo {author} {\bibfnamefont {Muneto}\ \bibnamefont
  {Nitta}}, \bibinfo {author} {\bibfnamefont {Hiromitsu}\ \bibnamefont
  {Takeuchi}}, \ and\ \bibinfo {author} {\bibfnamefont {Makoto}\ \bibnamefont
  {Tsubota}},\ }\bibfield  {title} {\enquote {\bibinfo {title} {{Interaction of
  half-quantized vortices in two-component Bose-Einstein condensates}},}\
  }\href {\doibase 10.1103/PhysRevA.83.063603} {\bibfield  {journal} {\bibinfo
  {journal} {Phys. Rev. A}\ }\textbf {\bibinfo {volume} {83}},\ \bibinfo
  {pages} {063603} (\bibinfo {year} {2011})}\BibitemShut {NoStop}%
\bibitem [{\citenamefont {Kasamatsu}\ \emph {et~al.}(2016)\citenamefont
  {Kasamatsu}, \citenamefont {Eto},\ and\ \citenamefont
  {Nitta}}]{vortex_interaction}%
  \BibitemOpen
  \bibfield  {author} {\bibinfo {author} {\bibfnamefont {Kenichi}\ \bibnamefont
  {Kasamatsu}}, \bibinfo {author} {\bibfnamefont {Minoru}\ \bibnamefont {Eto}},
  \ and\ \bibinfo {author} {\bibfnamefont {Muneto}\ \bibnamefont {Nitta}},\
  }\bibfield  {title} {\enquote {\bibinfo {title} {{Short-range intervortex
  interaction and interacting dynamics of half-quantized vortices in
  two-component Bose-Einstein condensates}},}\ }\href {\doibase
  10.1103/PhysRevA.93.013615} {\bibfield  {journal} {\bibinfo  {journal} {Phys.
  Rev. A}\ }\textbf {\bibinfo {volume} {93}},\ \bibinfo {pages} {013615}
  (\bibinfo {year} {2016})}\BibitemShut {NoStop}%
\bibitem [{\citenamefont {Stenger}\ \emph {et~al.}(1998)\citenamefont
  {Stenger}, \citenamefont {Inouye}, \citenamefont {Stamper-Kurn},
  \citenamefont {Miesner}, \citenamefont {Chikkatur},\ and\ \citenamefont
  {Ketterle}}]{Stenger}%
  \BibitemOpen
  \bibfield  {author} {\bibinfo {author} {\bibfnamefont {J.}~\bibnamefont
  {Stenger}}, \bibinfo {author} {\bibfnamefont {S.}~\bibnamefont {Inouye}},
  \bibinfo {author} {\bibfnamefont {D.~M.}\ \bibnamefont {Stamper-Kurn}},
  \bibinfo {author} {\bibfnamefont {H.~J.}\ \bibnamefont {Miesner}}, \bibinfo
  {author} {\bibfnamefont {A.~P.}\ \bibnamefont {Chikkatur}}, \ and\ \bibinfo
  {author} {\bibfnamefont {W.}~\bibnamefont {Ketterle}},\ }\bibfield  {title}
  {\enquote {\bibinfo {title} {{Spin domains in ground-state Bose--Einstein
  condensates}},}\ }\href {http://dx.doi.org/10.1038/24567} {\bibfield
  {journal} {\bibinfo  {journal} {Nature}\ }\textbf {\bibinfo {volume} {396}},\
  \bibinfo {pages} {345} (\bibinfo {year} {1998})}\BibitemShut {NoStop}%
\bibitem [{\citenamefont {Chang}\ \emph {et~al.}(2004)\citenamefont {Chang},
  \citenamefont {Hamley}, \citenamefont {Barrett}, \citenamefont {Sauer},
  \citenamefont {Fortier}, \citenamefont {Zhang}, \citenamefont {You},\ and\
  \citenamefont {Chapman}}]{Chang}%
  \BibitemOpen
  \bibfield  {author} {\bibinfo {author} {\bibfnamefont {M.-S.}\ \bibnamefont
  {Chang}}, \bibinfo {author} {\bibfnamefont {C.~D.}\ \bibnamefont {Hamley}},
  \bibinfo {author} {\bibfnamefont {M.~D.}\ \bibnamefont {Barrett}}, \bibinfo
  {author} {\bibfnamefont {J.~A.}\ \bibnamefont {Sauer}}, \bibinfo {author}
  {\bibfnamefont {K.~M.}\ \bibnamefont {Fortier}}, \bibinfo {author}
  {\bibfnamefont {W.}~\bibnamefont {Zhang}}, \bibinfo {author} {\bibfnamefont
  {L.}~\bibnamefont {You}}, \ and\ \bibinfo {author} {\bibfnamefont {M.~S.}\
  \bibnamefont {Chapman}},\ }\bibfield  {title} {\enquote {\bibinfo {title}
  {{Observation of Spinor Dynamics in Optically Trapped $^{87}\mathrm{Rb}$
  Bose-Einstein Condensates}},}\ }\href {\doibase
  10.1103/PhysRevLett.92.140403} {\bibfield  {journal} {\bibinfo  {journal}
  {Phys. Rev. Lett.}\ }\textbf {\bibinfo {volume} {92}},\ \bibinfo {pages}
  {140403} (\bibinfo {year} {2004})}\BibitemShut {NoStop}%
\bibitem [{\citenamefont {Stamper-Kurn}\ and\ \citenamefont
  {Ueda}(2013)}]{Stamper-Kurn}%
  \BibitemOpen
  \bibfield  {author} {\bibinfo {author} {\bibfnamefont {Dan~M.}\ \bibnamefont
  {Stamper-Kurn}}\ and\ \bibinfo {author} {\bibfnamefont {Masahito}\
  \bibnamefont {Ueda}},\ }\bibfield  {title} {\enquote {\bibinfo {title}
  {{Spinor Bose gases: Symmetries, magnetism, and quantum dynamics}},}\ }\href
  {\doibase 10.1103/RevModPhys.85.1191} {\bibfield  {journal} {\bibinfo
  {journal} {Rev. Mod. Phys.}\ }\textbf {\bibinfo {volume} {85}},\ \bibinfo
  {pages} {1191--1244} (\bibinfo {year} {2013})}\BibitemShut {NoStop}%
\bibitem [{\citenamefont {Kwon}\ \emph {et~al.}(2015)\citenamefont {Kwon},
  \citenamefont {Moon}, \citenamefont {Seo},\ and\ \citenamefont
  {Shin}}]{vortex_shedding}%
  \BibitemOpen
  \bibfield  {author} {\bibinfo {author} {\bibfnamefont {Woo~Jin}\ \bibnamefont
  {Kwon}}, \bibinfo {author} {\bibfnamefont {Geol}\ \bibnamefont {Moon}},
  \bibinfo {author} {\bibfnamefont {Sang~Won}\ \bibnamefont {Seo}}, \ and\
  \bibinfo {author} {\bibfnamefont {Y.}~\bibnamefont {Shin}},\ }\bibfield
  {title} {\enquote {\bibinfo {title} {{Critical velocity for vortex shedding
  in a Bose-Einstein condensate}},}\ }\href {\doibase
  10.1103/PhysRevA.91.053615} {\bibfield  {journal} {\bibinfo  {journal} {Phys.
  Rev. A}\ }\textbf {\bibinfo {volume} {91}},\ \bibinfo {pages} {053615}
  (\bibinfo {year} {2015})}\BibitemShut {NoStop}%
\bibitem [{\citenamefont {Petrov}\ \emph {et~al.}(2000)\citenamefont {Petrov},
  \citenamefont {Holzmann},\ and\ \citenamefont {Shlyapnikov}}]{Petrov}%
  \BibitemOpen
  \bibfield  {author} {\bibinfo {author} {\bibfnamefont {D.~S.}\ \bibnamefont
  {Petrov}}, \bibinfo {author} {\bibfnamefont {M.}~\bibnamefont {Holzmann}}, \
  and\ \bibinfo {author} {\bibfnamefont {G.~V.}\ \bibnamefont {Shlyapnikov}},\
  }\bibfield  {title} {\enquote {\bibinfo {title} {{Bose-Einstein Condensation
  in Quasi-2D Trapped Gases}},}\ }\href {\doibase 10.1103/PhysRevLett.84.2551}
  {\bibfield  {journal} {\bibinfo  {journal} {Phys. Rev. Lett.}\ }\textbf
  {\bibinfo {volume} {84}},\ \bibinfo {pages} {2551--2555} (\bibinfo {year}
  {2000})}\BibitemShut {NoStop}%
\bibitem [{\citenamefont {Muruganandam}\ and\ \citenamefont
  {Adhikari}(2009)}]{GPfortran}%
  \BibitemOpen
  \bibfield  {author} {\bibinfo {author} {\bibfnamefont {Paulsamy}\
  \bibnamefont {Muruganandam}}\ and\ \bibinfo {author} {\bibfnamefont
  {Sadhan~K}\ \bibnamefont {Adhikari}},\ }\bibfield  {title} {\enquote
  {\bibinfo {title} {{Fortran programs for the time-dependent Gross--Pitaevskii
  equation in a fully anisotropic trap}},}\ }\href {\doibase
  10.1016/j.cpc.2009.04.015} {\bibfield  {journal} {\bibinfo  {journal}
  {Computer Physics Communications}\ }\textbf {\bibinfo {volume} {180}},\
  \bibinfo {pages} {1888--1912} (\bibinfo {year} {2009})}\BibitemShut {NoStop}%
\bibitem [{\citenamefont {Bao}\ and\ \citenamefont
  {Lim}(2008)}]{GP_numerical_methods_spin_1}%
  \BibitemOpen
  \bibfield  {author} {\bibinfo {author} {\bibfnamefont {Weizhu}\ \bibnamefont
  {Bao}}\ and\ \bibinfo {author} {\bibfnamefont {Fong~Yin}\ \bibnamefont
  {Lim}},\ }\bibfield  {title} {\enquote {\bibinfo {title} {{Computing Ground
  States of Spin-1 Bose-Einstein Condensates by the Normalized Gradient
  Flow}},}\ }\href {\doibase 10.1137/070698488} {\bibfield  {journal} {\bibinfo
   {journal} {SIAM Journal on Scientific Computing}\ }\textbf {\bibinfo
  {volume} {30}},\ \bibinfo {pages} {1925--1948} (\bibinfo {year}
  {2008})}\BibitemShut {NoStop}%
\bibitem [{\citenamefont {Bao}\ and\ \citenamefont
  {Wang}(2007)}]{GP_numerical_spin_1_method_derivations}%
  \BibitemOpen
  \bibfield  {author} {\bibinfo {author} {\bibfnamefont {Weizhu}\ \bibnamefont
  {Bao}}\ and\ \bibinfo {author} {\bibfnamefont {Hanquan}\ \bibnamefont
  {Wang}},\ }\bibfield  {title} {\enquote {\bibinfo {title} {{A Mass and
  Magnetization Conservative and Energy-Diminishing Numerical Method for
  Computing Ground State of Spin-1 Bose-Einstein Condensates}},}\ }\href
  {\doibase 10.1137/070681624} {\bibfield  {journal} {\bibinfo  {journal} {SIAM
  Journal on Numerical Analysis}\ }\textbf {\bibinfo {volume} {45}},\ \bibinfo
  {pages} {2177--2200} (\bibinfo {year} {2007})}\BibitemShut {NoStop}%
\bibitem [{\citenamefont {{Gilbert Strang}}(1968)}]{Strang}%
  \BibitemOpen
  \bibfield  {author} {\bibinfo {author} {\bibnamefont {{Gilbert Strang}}},\
  }\bibfield  {title} {\enquote {\bibinfo {title} {{On the Construction and
  Comparison of Difference Schemes}},}\ }\href {\doibase 10.1137/0705041}
  {\bibfield  {journal} {\bibinfo  {journal} {SIAM Journal on Numerical
  Analysis}\ }\textbf {\bibinfo {volume} {5}},\ \bibinfo {pages} {506--517}
  (\bibinfo {year} {1968})},\ \Eprint
  {http://arxiv.org/abs/https://doi.org/10.1137/0705041}
  {https://doi.org/10.1137/0705041} \BibitemShut {NoStop}%
\bibitem [{\citenamefont {Vudragovi{\'c}}\ \emph {et~al.}(2012)\citenamefont
  {Vudragovi{\'c}}, \citenamefont {Vidanovi{\'c}}, \citenamefont {Bala{\v{z}}},
  \citenamefont {Muruganandam},\ and\ \citenamefont {Adhikari}}]{GPC}%
  \BibitemOpen
  \bibfield  {author} {\bibinfo {author} {\bibfnamefont {Du{\v{s}}an}\
  \bibnamefont {Vudragovi{\'c}}}, \bibinfo {author} {\bibfnamefont {Ivana}\
  \bibnamefont {Vidanovi{\'c}}}, \bibinfo {author} {\bibfnamefont {Antun}\
  \bibnamefont {Bala{\v{z}}}}, \bibinfo {author} {\bibfnamefont {Paulsamy}\
  \bibnamefont {Muruganandam}}, \ and\ \bibinfo {author} {\bibfnamefont
  {Sadhan~K}\ \bibnamefont {Adhikari}},\ }\bibfield  {title} {\enquote
  {\bibinfo {title} {{C programs for solving the time-dependent
  Gross--Pitaevskii equation in a fully anisotropic trap}},}\ }\href {\doibase
  10.1016/j.cpc.2012.03.022} {\bibfield  {journal} {\bibinfo  {journal}
  {Computer Physics Communications}\ }\textbf {\bibinfo {volume} {183}},\
  \bibinfo {pages} {2021--2025} (\bibinfo {year} {2012})}\BibitemShut {NoStop}%
\bibitem [{\citenamefont {Lovegrove}\ \emph {et~al.}(2012)\citenamefont
  {Lovegrove}, \citenamefont {Borgh},\ and\ \citenamefont
  {Ruostekoski}}]{SQV_energetic_stability_spin_1}%
  \BibitemOpen
  \bibfield  {author} {\bibinfo {author} {\bibfnamefont {Justin}\ \bibnamefont
  {Lovegrove}}, \bibinfo {author} {\bibfnamefont {Magnus~O.}\ \bibnamefont
  {Borgh}}, \ and\ \bibinfo {author} {\bibfnamefont {Janne}\ \bibnamefont
  {Ruostekoski}},\ }\bibfield  {title} {\enquote {\bibinfo {title}
  {{Energetically stable singular vortex cores in an atomic spin-1
  Bose-Einstein condensate}},}\ }\href {\doibase 10.1103/PhysRevA.86.013613}
  {\bibfield  {journal} {\bibinfo  {journal} {Phys. Rev. A}\ }\textbf {\bibinfo
  {volume} {86}},\ \bibinfo {pages} {013613} (\bibinfo {year}
  {2012})}\BibitemShut {NoStop}%
\bibitem [{Note1()}]{Note1}%
  \BibitemOpen
  \bibinfo {note} {We note that the couplings $g_{12},g_1,g_2$ defined in~\cite
  {EtoInteraction} are in our notation $g_{12}=c_0 - c_2$ and $g_1 = g_2 = c_0
  + c_2$.}\BibitemShut {Stop}%
\bibitem [{Note2()}]{Note2}%
  \BibitemOpen
  \bibinfo {note} {For SQVs, $Q_s = 0$ and Eq.~\protect \textup {\hbox
  {\mathsurround \z@ \protect \normalfont (\ignorespaces \ref
  {meanprdiff_AF}\unskip \@@italiccorr )}} becomes symmetric for $\protect
  \mathaccentV {tilde}07E{A}_{\pm 1}$. Thus, $\protect \mathaccentV
  {tilde}07E{A}_{1} = \protect \mathaccentV {tilde}07E{A}_{-1}$, which renders
  the spin interaction energy part in Eq.~\protect \textup {\hbox
  {\mathsurround \z@ \protect \normalfont (\ignorespaces \ref
  {energy_AF}\unskip \@@italiccorr )}} zero. Hence, $E_s$ does not depend on
  $c_2$. With no vortex in the AF phase, $Q_n = Q_s = 0$ and it is easily shown
  that, according to Eqs.~\protect \textup {\hbox {\mathsurround \z@ \protect
  \normalfont (\ignorespaces \ref {meanprdiff_AF}\unskip \@@italiccorr )}}
  and~\protect \textup {\hbox {\mathsurround \z@ \protect \normalfont
  (\ignorespaces \ref {energy_AF}\unskip \@@italiccorr )}}, $E_{\protect
  \textrm {ref}}$ does not depend on $c_2$.}\BibitemShut {Stop}%
\bibitem [{\citenamefont {Papoular}\ \emph {et~al.}(2010)\citenamefont
  {Papoular}, \citenamefont {Shlyapnikov},\ and\ \citenamefont
  {Dalibard}}]{changing_a0_by_microwave}%
  \BibitemOpen
  \bibfield  {author} {\bibinfo {author} {\bibfnamefont {D.~J.}\ \bibnamefont
  {Papoular}}, \bibinfo {author} {\bibfnamefont {G.~V.}\ \bibnamefont
  {Shlyapnikov}}, \ and\ \bibinfo {author} {\bibfnamefont {J.}~\bibnamefont
  {Dalibard}},\ }\bibfield  {title} {\enquote {\bibinfo {title}
  {{Microwave-induced Fano-Feshbach resonances}},}\ }\href {\doibase
  10.1103/PhysRevA.81.041603} {\bibfield  {journal} {\bibinfo  {journal} {Phys.
  Rev. A}\ }\textbf {\bibinfo {volume} {81}},\ \bibinfo {pages} {041603}
  (\bibinfo {year} {2010})}\BibitemShut {NoStop}%
\bibitem [{\citenamefont {Ding}\ \emph {et~al.}(2017)\citenamefont {Ding},
  \citenamefont {D'Incao},\ and\ \citenamefont {Greene}}]{a0_change_via_radio}%
  \BibitemOpen
  \bibfield  {author} {\bibinfo {author} {\bibfnamefont {Yijue}\ \bibnamefont
  {Ding}}, \bibinfo {author} {\bibfnamefont {Jos\'e~P.}\ \bibnamefont
  {D'Incao}}, \ and\ \bibinfo {author} {\bibfnamefont {Chris~H.}\ \bibnamefont
  {Greene}},\ }\bibfield  {title} {\enquote {\bibinfo {title} {{Effective
  control of cold collisions with radio-frequency fields}},}\ }\href {\doibase
  10.1103/PhysRevA.95.022709} {\bibfield  {journal} {\bibinfo  {journal} {Phys.
  Rev. A}\ }\textbf {\bibinfo {volume} {95}},\ \bibinfo {pages} {022709}
  (\bibinfo {year} {2017})}\BibitemShut {NoStop}%
\bibitem [{\citenamefont {{Cheianov}}\ and\ \citenamefont
  {{Chudnovskiy}}()}]{Cheianov}%
  \BibitemOpen
  \bibfield  {author} {\bibinfo {author} {\bibfnamefont {V.}~\bibnamefont
  {{Cheianov}}}\ and\ \bibinfo {author} {\bibfnamefont {A.~L.}\ \bibnamefont
  {{Chudnovskiy}}},\ }\bibfield  {title} {\enquote {\bibinfo {title}
  {{Microwave control of coupling parameters in spinor alkali condensates}},}\
  }\href@noop {} {\ }\Eprint {http://arxiv.org/abs/1705.01478}
  {arXiv:1705.01478 [cond-mat.quant-gas]} \BibitemShut {NoStop}%
\end{thebibliography}%

\end{document}